\documentclass[12pt]{iopart}

\expandafter\let\csname equation*\endcsname\relax
\expandafter\let\csname endequation*\endcsname\relax
\expandafter\let\csname align*\endcsname\relax
\expandafter\let\csname endalign*\endcsname\relax

\usepackage{amsmath, amssymb,amsfonts}%
\usepackage[utf8]{inputenc}
\usepackage[T1]{fontenc}
\usepackage{graphicx}%
\usepackage{multirow}%
\usepackage{amsthm}%
\usepackage{mathrsfs}%
\usepackage{xcolor}%
\usepackage{textcomp}%
\usepackage{manyfoot}%
\usepackage{booktabs}%
\usepackage{algorithm}%
\usepackage{algorithmicx}%
\usepackage{algpseudocode}%
\usepackage{listings}%
\usepackage{bbm}
\usepackage{comment}
\usepackage{calrsfs}
\usepackage{ytableau} 
\usepackage{cite}
\DeclareMathAlphabet{\pazocal}{OMS}{zplm}{m}{n}
\newcommand{\mcal}[1]{\pazocal{#1}}

\newcommand{\threenu}[3]{\left( \begin{array}{ccc} #1 & #2 & #3 \\ W_{#1} & W_{#2} & W_{#3} \end{array} \right)}
\newcommand{\genthreenu}[3]{\left( \begin{array}{ccc} #1 & \boldsymbol{#2} & #3 \\ W_{#1} & \boldsymbol{W}_{\boldsymbol{#2}} & W_{#3} \end{array} \right)}

\begin{document}

\title[Permutationally invariant processes in open multiqudit systems]{Permutationally invariant processes in open multiqudit systems}

\author{T Bastin$^{1}$\footnote{Author to whom any correspondence should be addressed.} and J Martin$^1$}

\address{$^1$ University of Liege (ULi\`ege), Institut de Physique Nucl\'eaire, Atomique et de Spectroscopie, CESAM, 4000 Li\`ege, Belgium}
\eads{\mailto{T.Bastin@uliege.be}, \mailto{jmartin@uliege.be}}

\begin{abstract}
 We establish the comprehensive theoretical framework for an exact description of the open system dynamics of permutationally invariant (PI) states in arbitrary $N$-qudit systems when this dynamics preserves the PI symmetry over time. Thanks to the powerful Schur-Weyl duality formalism, we unveil the internal links between the canonical time-local Lindblad-like master equation and the Markovian or non-Markovian dynamics of each permutationally-invariant degree of freedom (Schur subspaces). Our approach does not require one to compute the Schur transform as it operates directly within the restricted PI operator subspace of the Liouville space, whose dimension only scales polynomially with the number of qudits. We introduce the concept of $3\nu$-symbol matrix, where $\nu$ here denotes an integer partition, that proves to be very useful in this context.
\end{abstract}


\flushbottom
\maketitle

\thispagestyle{empty}

\section{Introduction}
\subsection{Background}
The ability to efficiently simulate the dynamics of noisy many-body quantum systems such as Noisy Intermediate-Scale Quantum (NISQ) devices is nowadays of primary importance, e.g., in order to assess whether they can offer a quantum advantage. To this end, it is necessary to solve a many-body master equation for the density matrix, which is intrinsically more complex than the Schr\"odinger equation and involves a number of variables that increases very unfavourably with the number of levels of the constituents (qudits). It is often required to go beyond Lindblad master equations since they only represent a simplified model that does not fully account for the non-Markovian nature of realistic environments where memory effects enter into play, such as spin-bath interactions in superconducting qubits (see, e.g., Ref.~\cite{Odeh2025}). In quantum science and technologies, while more delicate to handle, multilevel quantum systems have proven to offer several advantages over conventional two-level entities (qubits)~\cite{Mar17, Erh20}. These include higher information capacity~\cite{Erh20, Wan20}, increased resistance to noise~\cite{Awschalom2021,Ecker2019}, greater security in quantum key distribution~\cite{Cerf2002, She2010, Hub13}, more powerful metrological schemes~\cite{Bou17, Shl18}, and an improved ability for closing the detection loophole in Bell experiments~\cite{Ver2010}, for error correction~\cite{Grassl2018}, or also for quantum machine learning tasks~\cite{Zhe23}. Several physical platforms can be used to obtain multiqudit systems. For example, light, with its multiphoton states, is primarily a multiqubit system where the state of a qubit is encoded in the polarisation of a photon or in two of its spatial modes~\cite{Kok2007}. It can also embody a multiqudit system by giving photons access to $d>2$ distinct temporal modes or frequency modes~\cite{Reimer2019}, or by structuring light to confer orbital angular momentum to photons~\cite{Zeilinger2018,Willner2021}. Alternatively, individual neutral atoms, which are now routinely cooled, trapped in optical lattices and tweezers and internally controlled by laser light, are being used as registers of qubits and qudits~\cite{Schrader2004,Henriet2020}. Trapped ions~\cite{Low2020,Ringbauer2022}, ultracold atomic mixtures~\cite{Kasper2022} (where qudits are encoded in the collective spin of a few atoms whose number can be varied), superconducting devices~\cite{Cervera-Lierta2022}, nitrogen-vacancy (NV) centers in diamond~\cite{Bradley2019}, or even molecules~\cite{Moreno-Pineda2018,Sawant2020} are other physical platforms commonly used in this context. When the multiqudit system is composed of identical though not necessarily indistinguishable qudits, a rich variety of \emph{collective} dynamical behaviors can emerge, such as superradiance~\cite{Lin2012,Norcia2016}, spin-squeezing~\cite{Norris2012}, or also dissipative phase transitions~\cite{Lee2014} to name just a few. In this context, it is therefore essential to find efficient methods to describe the dynamics of the system. Some authors have developed such methods when dissipation acts only collectively or individually, first for qubits~\cite{Cha08,Xu13,Shammah2018} and later also for qudits~\cite{Gegg2016a,Gegg2016b,Huybrechts2021}. These methods have been applied in various studies~\cite{Lee2014,Rei02,Kir17,Whi22,Suk23,Ver23}, in particular for the critical interpretation of experiments on spin-squeezing and other collective atomic phenomena~\cite{Bar10}, to quantify the impact of recoil and individual atomic decay processes of indistinguishable atoms on collective phenomena~\cite{Dam16,Zha18}, or to reveal unexpected dissipative phase transitions, test the validity of mean-field theory, and explore the impact of dephasing on superradiance transitions in various models~\cite{Huy20,Sha21}. Notably, this approach has been extended to study dissipative all-to-all connected qudit systems~\cite{Gegg2016a,Gegg2016b,Huybrechts2021}, confirming its utility across diverse quantum scenarios, including ab initio approaches to x-ray cavity QED~\cite{Len21}. All these methods were mainly developed to be numerically useful and do not exploit the powerful connection with group representation theory similarly as in Ref.~\cite{Yad2023} in a thermodynamical context. Here we fill this gap and establish the general theoretical framework for an exact description of permutationally invariant processes in open multiqudit systems for both Markovian or non-Markovian dynamics.

\subsection{Permutationally invariant processes}
\label{PIPsection}
Under fairly general conditions, the dynamics of an open quantum system can be described by a master equation of the form~\cite{mEq1,mEq2}
 \begin{equation}
    \label{genmasterEq}
    \frac{d}{dt} \hat\rho(t) = \frac{i}{\hbar}\big[\hat{\rho}(t),\hat{H}_S(t)\big] + \int_0^t \mcal{K}_{s,t}[\hat{\rho}(s)]ds,
 \end{equation}
  where $\hat{\rho}(t)$ is the system density operator, $\hat{H}_S(t)$ the system Hamiltonian, and $\mcal{K}_{s,t}$ is a linear map that models the effects of the environment on the system. The general master equation (\ref{genmasterEq}) can often be written in a time-local form \begin{equation}
  \label{localInTimeMasterEq}
      \frac{d}{dt}\hat{\rho}(t)=\mcal{L}(t)[\hat{\rho}(t)],
  \end{equation}
  where the so-called Liouvillian superoperator $\mcal{L}(t)$ acts on the Liouville space $\mathcal{L}(\mcal{H})$ (the space of linear operators on the system Hilbert space $\mcal{H}$) and is such that $\mcal{L}(t)[\hat{\rho}]$ is Hermitian and traceless for all density operators $\hat{\rho}$~\cite{Hall2014}. The Liouvillian $\mcal{L}(t)$ can always be cast in a canonical Lindblad-like form~\cite{Hall2014}
  \begin{equation}
      \label{LindbladLiouvillian}
      \mcal{L}(t) = \mcal{V}(t) + \mcal{D}(t),
  \end{equation}
  with
  \begin{equation}
\mcal{V}(t)[\hat{\rho}] = \frac{i}{\hbar}\big[\hat{\rho},\hat{H}(t)\big],\quad \mcal{D}(t)[\hat{\rho}] = \sum_k \gamma_k(t)\mcal{D}_{\hat{L}_k(t)}[\hat{\rho}],
\end{equation}
where the Hamiltonian $\hat{H}(t)$ may incorporate environment-induced corrections and the sum over $k$ that contains at most $\mathrm{dim}(\mcal{H})^2 - 1$ terms runs over so-called decoherence channels characterized with positive or negative decoherence rates $\gamma_k(t)$ and jump operators $\hat{L}_k(t)$. For all operators $\hat{L}$, the superoperator $\mcal{D}_{\hat{L}}$ reads
\begin{equation}
\label{dissipator}
\mcal{D}_{\hat{L}}[\hat{\rho}] = \hat{L}\hat{\rho} \hat{L}^\dagger - \frac{1}{2} \{\hat{L}^\dagger \hat{L}, \hat{\rho} \}.
\end{equation}
When all decoherence rates $\gamma_k(t)$ and jump operators $\hat{L}_k(t)$ are independent of time and $\gamma_k(t) > 0, \forall k$, Eq.~(\ref{localInTimeMasterEq}) reduces to the well-known memoryless Lindblad master equation~\cite{Lin76, Bre02}. In all other cases, it describes non-Markovian dynamics (see, e.g., Refs.~\cite{Schaller2008,Mozgunov2020,Tscherbul2015,Trushechkin2021,Nathan2020,Davidovi2022,Merkli2022}).

The Liouvillian action (\ref{LindbladLiouvillian}) is fully determined given the Hamiltonian $\hat{H}(t)$, the set of rates $\gamma_k(t)$ and jump operators $\hat{L}_k(t)$. It is denoted accordingly $\mcal{L}_{\hat{H}(t),\{(\gamma_k(t),\hat{L}_k(t))\}} \equiv \mcal{V}_{\hat{H}(t)} + \mcal{D}_{\{(\gamma_k(t),\hat{L}_k(t))\}}$ if explicit notation is required. In what follows and for the seek of conciseness, the explicit dependence in time is not written anymore and is considered as implicit.

For an $N$-qudit system, the state space $\mcal{H}$ identifies to $\mcal{H}_d^{\otimes N}$, with $\mcal{H}_d \simeq \mathbb{C}^{d}$ the individual qudit state space~\cite{footnote00}. It has dimension $d^N$ and scales exponentially with $N$. Endowed with the standard Hilbert-Schmidt scalar product, the Liouville space $\mathcal{L}(\mcal{H})$ is itself a Hilbert space of dimension $d^{2N}$. This renders the curse of dimensionality already severe for moderate number of qudits. This severity can be significantly downgraded if the system exhibits large symmetries that constrain its dynamics in a much smaller-dimensional subspace of the Liouville space. This is in particular the case for so-called \emph{permutationally-invariant} (PI) states $\hat{\rho}$~\cite{Tot09} as long as the Liouvillian $\mcal{L}$ preserves the PI symmetry over time. A PI operator $\hat{A}_\mathrm{PI}$ is an operator that satisfies $\hat{P}_\sigma \hat{A}_\mathrm{PI}\hat{P}_\sigma^{\dagger}=\hat{A}_\mathrm{PI} \Leftrightarrow [\hat{P}_\sigma, \hat{A}_\mathrm{PI}] = 0$ for all permutations $\sigma$ of 1, \ldots, $N$, where $\hat{P}_\sigma$ denotes the standard unitary permutation operator associated with $\sigma$ in $\mcal{H}$~\cite{footnote0}. The vector subspace of PI operators in $\mathcal{L}(\mcal{H})$ is the so-called \emph{commutant} $\mathcal{L}_{S_N}(\mcal{H})$ of the (unitary) representation $\sigma \mapsto \hat{P}_\sigma$ on $\mcal{H}$ of the symmetric group $S_N$~\cite{Okounkov}. The commutant contains the identity operator and is closed under multiplication of operators and Hermitian conjugation~\cite{footnote1}. A superoperator $\mcal{L}$ preserves the PI symmetry if the commutant $\mathcal{L}_{S_N}(\mcal{H})$ is $\mcal{L}$-invariant, i.e., if $\mcal{L}[\hat{A}_\mathrm{PI}]$ is a PI operator regardless of the PI operator $\hat{A}_\mathrm{PI}$. It is also important that such superoperators avoid contaminating the commutant from any non-PI components, i.e., that the orthogonal complement of the commutant be itself $\mcal{L}$-invariant, which is equivalent to having the commutant $\mathcal{L}_{S_N}(\mcal{H})$ both $\mcal{L}$- and $\mcal{L}^\dagger$-invariant~\cite{footnote1b}.

The natural class of superoperators that preserve the PI symmetry and avoid contamination from any non-PI components is given by superoperators $\mcal{L}$ that are themselves PI in the sense that $[\mcal{P}_\sigma,\mcal{L}] = 0$ for all permutations $\sigma$, with $\mcal{P}_\sigma$ the (unitary) superoperator of permutation $\mcal{P}_\sigma[\hat{A}] = \hat{P}_\sigma \hat{A} \hat{P}_\sigma^\dagger, \forall \hat{A} \in \mathcal{L}(\mcal{H})$. Indeed, in this case for all PI operators $\hat{A}_\mathrm{PI}$, $\mcal{P}_\sigma \mcal{L}[\hat{A}_\mathrm{PI}] = \mcal{L} \mcal{P}_\sigma[\hat{A}_\mathrm{PI}] = \mcal{L}[\hat{A}_\mathrm{PI}], \forall \sigma$, i.e., $\mcal{L}[\hat{A}_{\mathrm{PI}}]$ is PI~\cite{footnote1c} and the commutant $\mathcal{L}_{S_N}(\mcal{H})$ is $\mcal{L}$-invariant. It is also $\mcal{L}^\dagger$-invariant since the space of PI superoperators is closed under Hermitian conjugation.

The superoperators of permutation satisfy $\mcal{P}_\sigma[\hat{A}\hat{B}] = \mcal{P}_\sigma[\hat{A}]\mcal{P}_\sigma[\hat{B}]$ and $\mcal{P}_\sigma[\hat{A}^\dagger] = \mcal{P}_\sigma[\hat{A}]^\dagger$~\cite{footnote2}. This implies interestingly that Liouvillians of the form of Eq.~(\ref{LindbladLiouvillian}) obey $\mcal{P}_\sigma \mcal{L}_{\hat{H},\{(\gamma_k,\hat{L}_k)\}} = \mcal{L}_{\mcal{P}_\sigma[\hat{H}],\{(\gamma_k,\mcal{P}_\sigma[\hat{L}_k])\}} \mcal{P}_\sigma$. If the Hamiltonian $\hat{H}$ is PI as well as the set $\{(\gamma_k,\hat{L}_k)\}$ of rates and jump operators \emph{as a whole}, i.e.,  $\{(\gamma_k,\mcal{P}_\sigma[\hat{L}_k])\} = \{(\gamma_k,\hat{L}_k)\}, \forall \sigma$ (which does not require to have individually $(\gamma_k,\mcal{P}_\sigma[\hat{L}_k]) = (\gamma_k,\hat{L}_k), \forall k, \sigma$), the Liouvillian $\mcal{L}_{\hat{H},\{(\gamma_k,\hat{L}_k)\}}$ is a PI superoperator. This is typically the case when the decoherence channels are composed of identical \emph{local} jump operators $\hat{\ell}^{(n)}$ associated with a unique local decoherence rate $\gamma_{\mathrm{loc}}$, $\forall n = 1, \ldots, N$, and/or a \emph{collective} jump operator $\hat{L}_c = \sum_{n=1}^N \hat{L}^{(n)}$ associated with its own decoherence rate $\gamma_c$, where the superscript $(n)$ denotes the specific qudit the local operator acts on~\cite{footnote3}. If both contributions are present, the superoperator $\mcal{D}$ contains a local and a collective part: $\mcal{D} = \mcal{D}^{(\mathrm{loc})}_{\hat{\ell}} + \mcal{D}^{(\mathrm{col})}_{\hat{L}}$, with \begin{equation}
\mcal{D}^{(\mathrm{loc})}_{\hat{\ell}} = \gamma_{\mathrm{loc}} \sum_{n=1}^N \mcal{D}_{\hat{\ell}^{(n)}}, \quad \mcal{D}^{(\mathrm{col})}_{\hat{L}} = \gamma_c \mcal{D}_{\hat{L}_c}.
\end{equation}

More general PI superoperators $\mcal{D}$ can also be envisaged with, for instance, identical two-particle jump operators $\hat{\ell}_2^{(n,m)}$, $\forall n < m = 1, \ldots, N$ associated with a unique decoherence rate $\gamma_2$, and/or a collective two-particle jump operator $\hat{L}_{2,c} = \sum_{n < m} \hat{L}_2^{(n,m)}$ associated with a decoherence rate $\gamma_{2,c}$, where $(n,m)$ denotes the particle pair the two-particle operators $\hat{\ell}_2$ and $\hat{L}_2$ act on. Strictly generally we can even consider identical $p$-particle ($p \leq N$) jump operators $\hat{\ell}_p^{(n_1,\ldots, n_p)}$ associated to a unique decoherence rate $\gamma_p$, $\forall n_1 < \cdots < n_p$, and also a collective $p$-particle jump operator $\hat{L}_{p,c} = \sum_{n_1 < \cdots < n_p} \hat{L}_p^{(n_1,\ldots, n_p)}$ with a decoherence rate $\gamma_{p,c}$, where $(n_1,\ldots, n_p)$ denotes the particle $p$-uple the $p$-particle operators $\hat{\ell}_p$ and $\hat{L}_p$ act on.

The commutant $\mathcal{L}_{S_N}(\mcal{H})$ is nothing but the symmetric subspace of the Liouville space $\mathcal{L}(\mcal{H})$~\cite{footnote3b}. Its dimension is thus equal to $\binom{N+d^2-1}{N}$~\cite{Okounkov} and scales only polynomially with $N$ in $\mcal{O}(N^{d^2-1})$ instead of exponentially as for the global Liouville space $\mathcal{L}(\mcal{H})$. This changes drastically the complexity class of PI systems for which large $N$ studies should remain more accessible within classical computational ressources. In this context, it is therefore highly desirable to develop tools that allow one to restrict the master equation treatment in the sole commutant subspace. This requires identifying a natural orthonormal basis of operators in $\mathcal{L}_{S_N}(\mcal{H})$ onto which the master equation can be projected and having explicit expressions of the matrix elements. This was specifically done in \cite{Cha08} for qubit systems ($d = 2$). For $d > 2$, nothing similar is identified, and we fill this gap in this work with the help of the powerful formalism of Schur-Weyl duality~(see, e.g., Refs.~\cite{Okounkov, Bacon06, Wallach}). The theory is established for arbitrary $d$ and the results for $d = 2$ are recovered as a special case.

\section{Results}
\subsection{Structure of the commutant $\mathcal{L}_{S_N}(\mcal{H})$}
\label{comSection}
The $N$-qudit state space $\mcal{H} = \mcal{H}_d^{\otimes N}$ is a natural representation space for both the symmetric group $S_N$ and the general linear group $GL(d) \equiv GL(d,\mathbb{C})$ of $d \times d$ invertible complex matrices (including its subgroup $U(d)$ of $d \times d$ unitary matrices). The standard representation operator for $\sigma \in S_N$ is the unitary permutation operator $\hat{P}_\sigma$ and for $A \in GL(d)$ the tensor product operator $\hat{A}^{\otimes N}$, with $\hat{A}$ the invertible local operator of representation matrix $A$ in the single-qudit basis. Both operators $\hat{P}_\sigma$ and $\hat{A}^{\otimes N}$ commute, so that the product operators $\hat{P}_\sigma\hat{A}^{\otimes N}$ define a representation of the direct product group $S_N \times GL(d)$ on $\mcal{H}$. For $N,d > 1$, $\mcal{H}$ is a reducible representation space for both the symmetric and the general linear group, as well as for the direct product group. As a consequence of the Schur-Weyl duality, the state space $\mcal{H}$ can be decomposed into irreducible subrepresentations of both $S_N$ and $GL(d)$, according to the multiplicity-free decomposition of the direct product group representation $\mcal{H} \simeq \oplus_{\nu \vdash (N,d)} \mcal{S}^\nu \otimes \mcal{U}^\nu(d)$, where the direct sum runs over all partitions $\nu$ of $N$ of at most $d$ parts, and $\mcal{S}^\nu$ and $\mcal{U}^\nu(d)$ denote the unitary irreducible representations (irreps) of $S_N$ and $GL(d)$ associated to $\nu$, respectively~\cite{footnoteSnuUnud}. The restriction of $\mcal{U}^\nu(d)$ onto the subgroup $U(d)$ is also irreducible and can be denoted similarly. We have on the one side $\mcal{H} \simeq \oplus_{\nu \vdash (N,d)} {\mcal{S}^\nu}^{\oplus \mathrm{dim}\, \mcal{U}^\nu(d)}$, and on the other side $\mcal{H} \simeq \oplus_{\nu \vdash (N,d)} {\mcal{U}^\nu(d)}^{\oplus \mathrm{dim}\, \mcal{S}^\nu}$. In this context, a natural basis in the state space $\mcal{H}$ is the orthonormal so-called \emph{Schur} basis~\cite{Zhe23, Bacon06} $\{|\nu, T_\nu, W_\nu\rangle, \forall \nu \vdash (N,d), T_\nu \in \mcal{T}_\nu, W_\nu \in \mcal{W}_\nu\}$, with $\mcal{T}_\nu$ the set of all standard Young Tableaux (SYT) $T_\nu$ of shape $\nu$ and $\mcal{W}_\nu$ the set of all semistandard Young Tableaux (SSYT) [also called standard Weyl Tableaux (SWT)] $W_\nu$ of shape $\nu$ and of content among $0, \ldots, d-1$. The cardinalities of the sets $\mcal{T}_\nu$ and $\mcal{W}_\nu$ are $f^\nu = \mathrm{dim}\,\mcal{S}^\nu$~\cite{footnotefnu} and $f^\nu(d) = \mathrm{dim}\,\mcal{U}^\nu(d)$~\cite{footnotefnud}, respectively. The Schur basis vectors $|\nu,T_\nu,W_\nu\rangle$ belong each to a well defined chain of irreps of both subgroup chains $S_N, S_{N-1}, \ldots, S_1$, and $U(d), U(d-1), \ldots, U(1)$. The two chains of irreps are encoded in the SYT $T_\nu$ for the symmetric group and in the SWT $W_\nu$ for the unitary group~\cite{footnoteTnuWnud}. For all $\nu \vdash (N,d)$ and $W_\nu \in \mcal{W}_\nu$, $\mcal{H}_\nu(W_\nu) \equiv \mathrm{span}\{|\nu, T_\nu, W_\nu\rangle, \forall T_\nu \in \mcal{T}_\nu\}$ is an $\mcal{S}^\nu$-equivalent irrep subspace of the symmetric group $S_N$. For all $\nu \vdash (N,d)$ and $T_\nu \in \mcal{T}_\nu$, $\mcal{H}_\nu(T_\nu) \equiv \mathrm{span}\{|\nu, T_\nu, W_\nu\rangle, \forall W_\nu \in \mcal{W}_\nu\}$ is an $\mcal{U}^\nu(d)$-equivalent irrep subspace of $U(d)$. In each of these irrep subspaces, the orthonormal vectors $|\nu, T_\nu, W_\nu\rangle$ identify to the unique (up to global phases) so-called Gel'fand-Tsetlin (GT) basis vectors of the irrep with respect to either of the above-cited subgroup chains~\cite{Okounkov, Vilenkin}.

We can now prove that an operator basis in the commutant $\mathcal{L}_{S_N}(\mcal{H})$ is nicely given by the set of PI operators
\begin{equation}
    \hat{F}_\nu^{(W_\nu,W'_\nu)} = \overline{|\nu,W_\nu\rangle \langle \nu,W'_\nu|} \equiv \frac{1}{\sqrt{f^\nu}} \sum_{T_\nu \in \mcal{T}_\nu} |\nu,T_\nu,W_\nu\rangle \langle \nu, T_\nu,W'_\nu|, \label{FnuWnuWpnu}
\end{equation}
$\forall \nu \vdash (N,d), W_\nu, W'_\nu \in \mcal{W}_\nu$. Indeed, these operators are easily seen to be permutationally invariant~\cite{footnote4} and their action on the Schur basis states reads $\hat{F}_\nu^{(W_\nu,W'_\nu)} |\nu',T_{\nu'},\tilde{W}_{\nu'}\rangle = 0$ if $\nu' \neq \nu$ and
\begin{equation}
    \hat{F}_\nu^{(W_\nu,W'_\nu)} |\nu,T_{\nu},\tilde{W}_{\nu}\rangle = \frac{1}{\sqrt{f^\nu}} |\nu, T_\nu, W_\nu\rangle \delta_{\tilde{W}_{\nu},W'_\nu},
\end{equation}
with $\delta$ the Kronecker delta. Hence, their range and kernel are given by $\mathrm{ran}\, \hat{F}_\nu(W_\nu, W'_\nu) = \mcal{H}_\nu(W_\nu)$ and $\mathrm{ker}\, \hat{F}_\nu^{(W_\nu, W'_\nu)} = \mcal{H} \ominus \mcal{H}_\nu(W'_\nu)$, respectively. Each operator $\hat{F}_\nu^{(W_\nu,W'_\nu)}$ maps a specific $\mcal{S}^\nu$-equivalent irrep subspace onto an equivalent one: $\hat{F}_\nu^{(W_\nu, W'_\nu)} \mcal{H}_\nu(W'_\nu) = \mcal{H}_\nu(W_\nu)$. All this makes the set of operators (\ref{FnuWnuWpnu}) an operator basis in the commutant $\mathcal{L}_{S_N}(\mcal{H})$~\cite{Okounkov} (as a corollary, $\mathrm{dim}\,\mathcal{L}_{S_N}(\mcal{H})$ can also be written $\sum_{\nu \vdash(N,d)}f^\nu(d)^2$~[see~\ref{dimCommutantApp}]). In addition, with respect to the standard Hilbert-Schmidt scalar product between any two linear operators, this basis is orthonormal:
\begin{equation}
    \mathrm{Tr}\left(\hat{F}_\nu^{(W_\nu,W'_\nu)\dagger} \hat{F}_{\nu'}^{(\tilde{W}_{\nu'},\tilde{W}'_{\nu'})}\right) = \delta_{\nu,\nu'} \delta_{W_\nu,\tilde{W}_{\nu'}} \delta_{W'_\nu,\tilde{W}'_{\nu'}}.
    \label{orthonormality}
\end{equation}
It follows that any PI operator $\hat{A}_\mathrm{PI}$ admits the expansion
\begin{equation}
    \hat{A}_\mathrm{PI} = \sum_{\nu \vdash(N,d)}\sum_{W_\nu, W'_\nu \in \mcal{W}_\nu} A_{\nu, W_\nu, W'_\nu} \hat{F}_\nu^{(W_\nu,W'_\nu)},
\end{equation}
with components $A_{\nu, W_\nu, W'_\nu} \equiv (\hat{A}_\mathrm{PI})_{\nu, W_\nu, W'_\nu}$ given by
\begin{equation}
    A_{\nu, W_\nu, W'_\nu} = \mathrm{Tr}(\hat{F}_\nu^{(W_\nu,W'_\nu)\dagger} \hat{A}_\mathrm{PI}) = \frac{1}{\sqrt{f^\nu}} \sum_{T_\nu \in \mcal{T}_\nu} \langle \nu, T_\nu, W_\nu|\hat{A}_\mathrm{PI}|\nu, T_\nu,W'_\nu\rangle.
\end{equation}
The matrix representation of such operators is block diagonal in the Schur basis \{$|\nu, T_\nu, W_\nu\rangle$\} if the basis vectors are sorted first by $\nu$, then by SYT $T_\nu$, and finally by SWT $W_\nu$, i.e., by vector subspaces $\mcal{H}_\nu(T_\nu), \forall \nu, T_\nu$. Blocks are of dimension $f^\nu(d) \times f^\nu(d)$ and only depend on $\nu$, but not on $T_\nu$, so that the representation matrix $A_\mathrm{PI}$ exhibits a double block-diagonal structure, with large ``$\nu$-blocks'', themselves composed of $f^\nu$ identical blocks $A(\nu)$: $A_\mathrm{PI} = \oplus_\nu A(\nu)^{\oplus f^\nu}$. The elements of a block $A(\nu)$ read $A(\nu)_{W_\nu, W'_\nu} = A_{\nu, W_\nu, W'_\nu}/\sqrt{f^\nu}$. In this representation, the trace of the PI operator $\hat{A}_\mathrm{PI}$ reads
\begin{equation}
    \mathrm{Tr}(\hat{A}_\mathrm{PI}) = \sum_{\nu \vdash (N,d)}\sum_{W_\nu \in \mcal{W}_\nu} \sqrt{f^\nu} A_{\nu,W_\nu,W_\nu}.
\end{equation}

The commutant can be decomposed into the direct sum of orthogonal operator subspaces $\mathcal{L}_\nu(\mcal{H}) \equiv \mathrm{span}\{\hat{F}_\nu^{(W_\nu,W'_\nu)}, \forall W_\nu, W'_\nu \in \mcal{W}_\nu\}$:
\begin{equation}
\mathcal{L}_{S_N}(\mcal{H}) = \bigoplus_{\nu \vdash (N,d)} \mathcal{L}_\nu(\mcal{H})
\end{equation}
and a PI operator that specifically belongs to a subspace $\mathcal{L}_\nu(\mcal{H})$ is hereafter referenced as a $\nu$-type operator.

The PI orthonormal basis operators $\hat{F}_\nu^{(W_\nu,W'_\nu)}$ are mutually Hermitian conjugate: $\hat{F}_\nu^{(W_\nu,W'_\nu)\dagger} = \hat{F}_\nu^{(W'_\nu,W_\nu)}$~\cite{footnote4b}. They fulfill the \emph{multiplication rule}~\cite{footnote5}
\begin{align}
\hat{F}_\nu^{(W_\nu,W'_\nu)} \hat{F}_{\nu'}^{(\tilde{W}_{\nu'},\tilde{W}'_{\nu'})} = \frac{1}{\sqrt{f^\nu}} \delta_{\nu,\nu'} \delta_{W'_\nu,\tilde{W}_{\nu'}} \hat{F}_\nu^{(W_\nu,\tilde{W}'_{\nu'})}.
\end{align}
As a result the components of a PI operator Hermitian conjugate are given by $(\hat{A}_\mathrm{PI}^\dagger)_{\nu, W_\nu, W'_\nu} = A_{\nu, W'_\nu, W_\nu}^*$ and those of a PI operator product read
\begin{equation}
    (\hat{A}_\mathrm{PI} \hat{B}_\mathrm{PI})_{\nu, W_\nu, W'_\nu} = \frac{1}{\sqrt{f^\nu}} \sum_{\tilde{W}_\nu \in \mcal{W}_\nu} A_{\nu, W_\nu, \tilde{W}_\nu} B_{\nu, \tilde{W}_\nu, W'_\nu}.
\end{equation}
Hence, not only the operator subspaces $\mathcal{L}_\nu(\mcal{H})$ are closed under multiplication of operators and Hermitian conjugation~\cite{footnote6}, but also left- or right-multiplying a PI operator with a $\nu$-type operator again yields a $\nu$-type operator. More generally, the product of any number of PI operators is of $\nu$-type as soon as so is one of the operator. Finally, we have the closure relation
\begin{equation}
\sum_{\nu \vdash (N,d)} \sum_{W_\nu \in \mcal{W}_\nu} \sqrt{f^\nu} \; \overline{|\nu,W_\nu\rangle \langle \nu,W_\nu|} = \hat{\mathbbm{1}}.
\label{closure}
\end{equation}

\subsection{Master equation, 3$\nu$ symbols, and general Identity}
\label{genIdSection}
Let the $N$-qudit system be initially in a PI state $\hat{\rho}_\mathrm{PI}(0)$ with a time evolution governed by a PI Liouvillian $\mcal{L}$. In this case, the system state is constrained within the commutant $\mathcal{L}_{S_N}(\mcal{H})$ and the PI operators $\hat{F}_\nu^{(W_\nu,W'_\nu)}$ provide us with a natural orthonormal operator basis onto which the master equation can be projected. We have, $\forall \lambda \vdash (N,d), W_\lambda, W'_\lambda \in \mcal{W}_\lambda$,
\begin{equation}
\label{me}
    \dot{\rho}_{\lambda,W_\lambda,W'_\lambda} = \sum_{\nu \vdash (N,d)}\sum_{W_\nu,W'_\nu \in \mcal{W}_\nu} \mcal{L}_{\lambda,W_\lambda,W'_\lambda; \nu, W_\nu,W'_\nu} \, \rho_{\nu,W_\nu,W'_\nu},
\end{equation}
where for any superoperator $\mcal{O}$
\begin{equation}
\mcal{O}_{\lambda,W_\lambda,W'_\lambda; \nu, W_\nu,W'_\nu} \equiv \mathrm{Tr}\left(\hat{F}_\lambda^{(W_\lambda,W'_\lambda)\dagger} \mcal{O}[\hat{F}_\nu^{(W_\nu,W'_\nu)}]\right)
\end{equation}
is the component of operator $\mcal{O}[\hat{F}_\nu^{(W_\nu,W'_\nu)}]$ along the commutant basis operator $\hat{F}_\lambda^{(W_\lambda,W'_\lambda)}$.

For a standard PI Liouvillian $\mcal{L} = \mcal{V}_{\hat{H}_c} + \mcal{D}^{(\mathrm{loc})}_{\hat{\ell}} + \mcal{D}^{(\mathrm{col})}_{\hat{L}}$, with $\hat{H}_c = \sum_n \hat{H}^{(n)}$, where $\hat{H}$ is a local (single particle) Hamiltonian and $\hat{\ell}$ and $\hat{L}$ are single-particle jump operators (more general PI Liouvillians with $p$-particle terms in either coherent or dissipative parts are discussed in~\ref{generalpPILindApp}), both operators $\mcal{V}_{\hat{H}_c}[\hat{F}_\nu^{(W_\nu,W'_\nu)}]$ and $\mcal{D}^{(\mathrm{col})}_{\hat{L}}[\hat{F}_\nu^{(W_\nu,W'_\nu)}]$ are of $\nu$-type because they are composed of products of PI operators with the $\nu$-type operator $\hat{F}_\nu^{(W_\nu,W'_\nu)}$. Their expansion in the commutant operator basis follows straightforwardly provided this expansion is explicitly known for each of the involved PI operators. More generally all Liouvillian terms can be expressed with the help of the superoperators $\mcal{K}_{\hat{X},\hat{Y}}$ ($\hat{X}, \hat{Y}$ are any two local operators) defined as
\begin{equation}
    \label{KXYdef}
    \mcal{K}_{\hat{X},\hat{Y}}[\hat{A}] = \sum_{n = 1}^N \hat{X}^{(n)} \hat{A} \hat{Y}^{(n)\dagger}, \quad \forall \hat{A} \in \mathcal{L}(\mcal{H}).
\end{equation}
Indeed, $\mcal{V}_{\hat{H}_c} = (i/\hbar) (\mcal{K}_{\hat{\mathbbm{1}},\hat{H}} - \mcal{K}_{\hat{H},\hat{\mathbbm{1}}})$,
\begin{equation}
\mcal{D}_{\hat{\ell}}^{(\mathrm{loc})} = \gamma_{\mathrm{loc}}(\mcal{K}_{\hat{\ell},\hat{\ell}} - \frac{1}{2} \mcal{K}_{\hat{\ell}^\dagger \hat{\ell},\hat{\mathbbm{1}}} - \frac{1}{2} \mcal{K}_{\hat{\mathbbm{1}},\hat{\ell}^\dagger \hat{\ell}}),
\end{equation}
and
\begin{equation}
\mcal{D}_{\hat{L}}^{(\mathrm{col})}[\hat{\rho}] = \gamma_c (\hat{L}_c \mcal{K}_{\hat{\mathbbm{1}},\hat{L}}[\hat{\rho}] - \frac{1}{2} \hat{L}_c^\dagger \mcal{K}_{\hat{L},\hat{\mathbbm{1}}}[\hat{\rho}] - \frac{1}{2} \mcal{K}_{\hat{\mathbbm{1}},\hat{L}}[\hat{\rho}] \hat{L}_c ),
\end{equation}
where $\hat{L}_c$ can similarly be written as $\mcal{K}_{\hat{L},\hat{\mathbbm{1}}}[\hat{\mathbbm{1}}]$. The superoperators $\mcal{K}_{\hat{X},\hat{Y}}$ are PI, so that $\mcal{K}_{\hat{X},\hat{Y}}[\hat{A}_\mathrm{PI}]$ is itself a PI operator for any PI operator $\hat{A}_\mathrm{PI}$. With respect to Hermitian conjugation, we have $\mcal{K}_{\hat{X},\hat{Y}}[\hat{A}]^\dagger = \mcal{K}_{\hat{Y},\hat{X}}[\hat{A}^\dagger]$ and $\mcal{K}_{\hat{X},\hat{Y}}^\dagger = \mcal{K}_{\hat{X}^\dagger,\hat{Y}^\dagger}$.

To get explicit expressions of the matrix elements $\mcal{L}_{\lambda,W_\lambda,W'_\lambda; \nu, W_\nu,W'_\nu}$, it is therefore enough to
have the expansion in the commutant operator basis of the PI operators $\mcal{K}_{\hat{X},\hat{Y}}[\hat{F}_\nu^{(W_\nu,W'_\nu)}]$, $\forall \hat{X},\hat{Y},\nu, W_\nu, W'_\nu$. Schur-Weyl duality formalism, trace invariance under cyclic permutations, and Clebsch-Gordan decomposition of tensorial products of unitary irreducible representations of the unitary group $U(d)$ allow one to obtain these expansions. To this aim, we denote for all $\nu \in \mathcal{P}_d$ (the set of partitions of at most $d$ parts) by $\nu^-$ [$\nu^+$] any partition $\in \mathcal{P}_d$ obtained by the removal [addition] of an inner [outer] corner of $\nu$~\cite{footnoteCorner}. The actions of removing [adding] an inner [outer] corner of a partition $\nu$ can be combined, so that $\nu^{-+}$ denotes any partition $\in \mathcal{P}_d$ obtained first by the removal of an inner corner of $\nu$, then by the addition of an outer corner of the resulting partition at first step.

For every $\nu_L, \nu, \nu_R \in \mathcal{P}_d$, $W_{\mu} \in \mcal{W}_{\mu}$ ($\mu = \nu_L, \nu, \nu_R$), we also introduce the $3\nu$ symbol \scalebox{0.6}{$\threenu{\nu_L}{\nu}{\nu_R}$} as being the square $d \times d$ matrix
 with entries
\begin{equation}
    \threenu{\nu_L}{\nu}{\nu_R}_{i,j} = \langle W_{\nu} , i | W_{\nu_L} \rangle \langle W_{\nu} , j | W_{\nu_R} \rangle, \qquad \forall i,j = 0, \ldots, d-1, \label{threenu}
\end{equation}
where $\langle W_{\nu} , i | W_{\nu_L} \rangle$ and $\langle W_{\nu} , j | W_{\nu_R} \rangle$ denote Clebsch-Gordan coefficients (CGC's) of the tensorial product $\mcal{U}^\nu(d) \otimes \mcal{U}^{(1)}(d)$ for the Gel'fand-Tsetlin bases (see~\ref{CGApp}). For all $\mu,\nu \in \mathcal{P}_d$, $W_\mu \in \mcal{W}_\mu$, $W_\nu \in \mcal{W}_\nu$, $k = 0, \ldots, d-1$, a CGC $\langle W_{\mu}, k | W_\nu \rangle$ is zero iff the following two conditions are not simultaneously satisfied (CGC selection rules): $\mu \in \{\nu^-\}$ and $W_\mu \in \mcal{W}_\mu^{(-k)}(W_\nu)$, where $\mcal{W}_\mu^{(\pm k)}(W_\nu)$ denotes the set of all SWT's $W_\mu$ of shape $\mu$, same content as $W_\nu$ $\pm$ one box $k$, and same Gel'fand-Tsetlin's pattern as that of $W_\nu$ $\pm$ one triangular shift pattern. The set $\mcal{W}_\mu^{(\pm k)}(W_\nu)$ is a subset of the set $\tilde{\mcal{W}}_\mu^{(\pm k)}(W_\nu)$ of \emph{all} SWT's of shape $\mu$ and same content as $W_\nu$ $\pm$ one box $k$. Its cardinality is at most $(d-1)!/k!$ (in particular, $1$ if $d = 2$ or $k = d-1$). 

It follows from the CGC selection rules that the $3\nu$-symbol matrix \scalebox{0.6}{$\threenu{\nu_L}{\nu}{\nu_R}$} is necessarily zero if the condition $\nu \in \{\nu_L^-\} \cap \{\nu_R^-\}$ (\emph{partition triangle selection rule}) is not satisfied. This condition can only be met if $\nu_L \in \{\nu_R^{-+}\}$ or equivalently $\nu_R \in \{\nu_L^{-+}\}$. We define the \emph{partition triangular delta} $\{\nu_L, \nu, \nu_R \}$ to be $1$ if the partition triangle selection rule is satisfied and $0$ otherwise. If $\{\nu_L, \nu, \nu_R \} = 1$, an individual element $i,j$ of the $3\nu$-symbol matrix is zero iff $W_{\nu} \notin \mcal{W}_\nu^{(-i)}(W_{\nu_L}) \cap \mcal{W}_\nu^{(-j)}(W_{\nu_R})$. The CGC's are real and so are the $3\nu$-symbol matrices. We thus have
\begin{equation}
\threenu{\nu_L}{\nu}{\nu_R} = \threenu{\nu_R}{\nu}{\nu_L}^T.
\end{equation}
The $3\nu$-symbol matrices obey the orthogonality relation [see Eq.~(\ref{orthogonality})]
\begin{equation}
    \sum_{W_\nu \in \mcal{W}_\nu} \mathrm{Tr}\left[\threenu{\nu_L}{\nu}{\nu_R}\right] = \{\nu_L,\nu,\nu_R\} \delta_{\nu_L, \nu_R}\delta_{W_{\nu_L},W_{\nu_R}}
\end{equation}
and they represent in the single-qudit basis $\{|i\rangle, i = 0, \ldots, d - 1 \}$ the single qudit operators
\begin{equation}
 \hat{g}_{\nu,W_\nu}^{(\nu_L,W_{\nu_L};\nu_R,W_{\nu_R})} = |\phi^{(\nu_L, W_{\nu_L})}_{\nu, W_\nu}\rangle \langle \phi^{(\nu_R, W_{\nu_R})}_{\nu, W_\nu}|,
\end{equation}
where we defined $\forall \mu,\nu \in \mathcal{P}_d$, $W_\mu \in \mcal{W}_\mu$, and $W_\nu \in \mcal{W}_\nu$, the \emph{unnormalized} single-qudit states $|\phi^{(\nu, W_{\nu})}_{\mu, W_\mu}\rangle = \sum_{i = 0}^{d-1} \langle W_{\mu} , i | W_{\nu} \rangle |i\rangle$. This sum contains at most one term since the Clebsch-Gordan coefficient $\langle W_{\mu} , i | W_{\nu} \rangle$ requires $W_{\mu} \in \mcal{W}^{(-i)}_\mu(\mcal{W}_\nu)$ to be nonzero and this can possibly only happen for a single index $i$. Hence, the $\hat{g}_{\nu,W_\nu}^{(\nu_L,W_{\nu_L};\nu_R,W_{\nu_R})}$ operator is either 0 or a multiple of the dyadic operator $|i\rangle \langle j|$ for $W_{\nu} \in \mcal{W}^{(-i)}_\nu(\mcal{W}_{\nu_L}) \cap \mcal{W}^{(-j)}_\nu(\mcal{W}_{\nu_R})$.

If $W_\mu \notin \mcal{W}_\mu$ for $\mu = \nu_L$, $\nu_R$, and/or $\nu$, the $3\nu$-symbol matrix \scalebox{0.6}{$\threenu{\nu_L}{\nu}{\nu_R}$} is not defined. However, it may be convenient to adopt the convention that it nevertheless exists and just identifies to the null matrix.

With this stated, we obtain the \emph{general Identity} (see proof in~\ref{generalIdentityApp})
\begin{equation}
\mcal{K}_{\hat{X},\hat{Y}}[\hat{F}_\nu^{(W_\nu,W'_\nu)}] = \sum_{\lambda \in \{ \nu^{-+}\}} \sum_{W_\lambda, W'_\lambda \in \mcal{W}_\lambda} K_{\hat{X},\hat{Y}}^{(\lambda, W_\lambda, W'_\lambda; \nu, W_\nu, W'_\nu)} \hat{F}_\lambda^{(W_\lambda,W'_\lambda)}, \label{genIdentity1}
\end{equation}
where
\begin{equation}
K_{\hat{X},\hat{Y}}^{(\lambda, W_\lambda, W'_\lambda; \nu, W_\nu, W'_\nu)} = \sum_{\mu \in \{\nu^-\} \cap \{\lambda^-\}} \sqrt{r^\mu_\nu r^\mu_\lambda} \mathrm{Tr}[\hat{g}^{(\lambda,W_\lambda;\nu,W_\nu)\dagger}_\mu \hat{X}] \mathrm{Tr}[\hat{g}^{(\lambda,W'_\lambda;\nu,W'_\nu)\dagger}_\mu \hat{Y}]^\ast, \label{coefGenIdentity}
\end{equation}
with $r^\mu_\nu \equiv N f^\mu/f^\nu$, $\forall \nu \vdash N, \mu \in \{\nu^-\}$, and $\hat{g}^{(\lambda,W_\lambda;\nu,W_\nu)}_\mu$ the single qudit operator
\begin{equation}
\hat{g}^{(\lambda,W_\lambda;\nu,W_\nu)}_\mu = \sum_{W_\mu \in \mcal{W}_\mu}  \hat{g}_{\mu,W_\mu}^{(\lambda,W_{\lambda};\nu,W_\nu)}.
\end{equation}
This operator vanishes if $\{\lambda,\mu,\nu\} = 0$. It satisfies $\hat{g}^{(\lambda,W_\lambda;\nu,W_\nu)\dagger}_\mu = \hat{g}^{(\nu,W_\nu;\lambda,W_\lambda)}_\mu$ and
\begin{equation}
\mathrm{Tr}[\hat{g}^{(\lambda,W_\lambda;\nu,W_\nu)}_\mu] = \{\lambda,\mu,\nu\}\delta_{\lambda, \nu}\delta_{W_{\lambda},W_{\nu}}.
\label{Trq}
\end{equation}
As a result, $\hat{\rho}^{(\nu,W_\nu)}_\mu \equiv \hat{g}^{(\nu,W_\nu;\nu,W_\nu)}_\mu$ is a trace~1 sum of projection operators, hence positive semidefinite, and represents a single qudit mixed state for every $\mu \in \{\nu^-\}$. The general Identity~(\ref{genIdentity1}) states equivalently that the matrix elements of the superoperator 
$\mcal{K}_{\hat{X},\hat{Y}}$ are given by
\begin{equation}
[\mcal{K}_{\hat{X},\hat{Y}}]_{\lambda, W_\lambda, W'_\lambda; \nu, W_\nu, W'_\nu} = K_{\hat{X},\hat{Y}}^{(\lambda, W_\lambda, W'_\lambda; \nu, W_\nu, W'_\nu)} \delta_{\lambda,\{\nu^{-+}\}},
\end{equation}
where we have added here the factor $\delta_{\lambda,\{\nu^{-+}\}}$ (1 if $\lambda \in \{\nu^{-+}\}$ and 0 otherwise) for an explicit reference on when the matrix elements are necessarily zero or not (this is superfluous since the partition triangle selection rule defined above implies $K_{\hat{X},\hat{Y}}^{(\lambda, W_\lambda, W'_\lambda; \nu, W_\nu, W'_\nu)} = 0$ if $\lambda \notin \{\nu^{-+}\}$). We could have equivalently written $\delta_{\nu,\{\lambda^{-+}\}}$.

Equation (\ref{genIdentity1}) generalizes to arbitrary multiqudit systems and local operators Identity~1 of Ref.~\cite{Cha08} that was developed in the specific context of multiqubit systems. The latter was obtained using an inductive approach non-extendable to multilevel systems. Here, a completely different approach based on the powerful Schur-Weyl duality formalism with newly introduced $3\nu$ symbols was followed to get the sought generalization to arbitrary $d$. 

Thanks to Eq.~(\ref{Trq}), the coefficients $K_{\hat{X},\hat{\mathbbm{1}}}^{(\lambda, W_\lambda, W'_\lambda; \lambda, \tilde{W}_\lambda, W'_\lambda)}$ are independent of $W'_\lambda$ and we can define
\begin{equation}
\label{CX}
K_{\hat{X}}^{(\lambda,W_\lambda, \tilde{W}_\lambda)} \equiv K_{\hat{X},\hat{\mathbbm{1}}}^{(\lambda, W_\lambda, W'_\lambda; \lambda, \tilde{W}_\lambda, W'_\lambda)} = \sum_{\mu \in \{\lambda^-\}} r^{\mu}_\lambda \mathrm{Tr}[\hat{g}^{(\lambda,W_\lambda;\lambda,\tilde{W}_\lambda)\dagger}_{\mu} \hat{X}].
\end{equation}
This yields $K_{\hat{X},\hat{\mathbbm{1}}}^{(\lambda, W_\lambda, W'_\lambda; \nu, W_\nu, W'_\nu)} = K_{\hat{X}}^{(\lambda,W_\lambda, W_\nu)} \delta_{\lambda, \nu} \delta_{W'_\lambda,W'_\nu}$ and subsequently~\cite{footnote7}
\begin{equation}
\label{genidentity1X1}
\mcal{K}_{\hat{X},\hat{\mathbbm{1}}}[\hat{F}_\nu^{(W_\nu,W'_\nu)}] = \sum_{\tilde{W}_\nu} K_{\hat{X}}^{(\nu, \tilde{W}_\nu, W_\nu)} \hat{F}_\nu^{(\tilde{W}_\nu,W'_\nu)}.
\end{equation}
Thanks to the closure relation (\ref{closure}), it follows that any collective operator $\hat{X}_c = \mcal{K}_{\hat{X},\hat{\mathbbm{1}}}[\hat{\mathbbm{1}}]$ can be written
\begin{equation}
    \hat{X}_c = \sum_{\nu \vdash(N,d)} \sum_{W_\nu, W'_\nu \in \mcal{W}_\nu} \sqrt{f^\nu} K_{\hat{X}}^{(\nu,W_\nu, W'_\nu)} \hat{F}_\nu^{(W_\nu,W'_\nu)}.
\end{equation}

For any local operators $\hat{X}$ and $\hat{Y}$, the coefficients (\ref{coefGenIdentity}) and (\ref{CX}) satisfy the symmetry relations
\begin{equation}
\label{symrelCXY}
K_{\hat{X},\hat{Y}}^{(\lambda,W'_\lambda,W_\lambda;\nu,W'_\nu,W_\nu)} = K_{\hat{Y},\hat{X}}^{(\lambda,W_\lambda,W'_\lambda;\nu,W_\nu,W'_\nu)*}, \qquad
K_{\hat{X}}^{(\lambda, \tilde{W}_\lambda,W_\lambda)} = K_{\hat{X}^\dagger}^{(\lambda, W_\lambda, \tilde{W}_\lambda)*}.
\end{equation}

The matrix elements $\mcal{L}_{\lambda, W_\lambda, W'_\lambda; \nu, W_\nu, W'_\nu}$ for the Liouvillian $\mcal{L} = \mcal{V}_{\hat{H}_c} + \mcal{D}^{(\mathrm{loc})}_{\hat{\ell}} + \mcal{D}^{(\mathrm{col})}_{\hat{L}}$ immediately follow from this formalism. We have
\begin{align}
    \mcal{L}_{\lambda, W_\lambda, W'_\lambda; \nu, W_\nu, W'_\nu} = {} & [\mcal{V}_{\hat{H}_c}]_{\lambda, W_\lambda, W'_\lambda; \nu, W_\nu, W'_\nu} + [\mcal{D}^{(\mathrm{loc})}_{\hat{\ell}}]_{\lambda, W_\lambda, W'_\lambda; \nu, W_\nu, W'_\nu} \\
    & + [\mcal{D}^{(\mathrm{col})}_{\hat{L}}]_{\lambda, W_\lambda, W'_\lambda; \nu, W_\nu, W'_\nu}.
    \nonumber
\end{align}

The commutator between any PI operator $\hat{A}_\mathrm{PI}$ and the basis operator $\hat{F}_\nu^{(W_\nu,W'_\nu)}$ follows straightforwardly from the commutant algebra multiplication rule. We have
\begin{equation}
    [\hat{F}_\nu^{(W_\nu,W'_\nu)},\hat{A}_\mathrm{PI}] = \frac{1}{\sqrt{f^\nu}} \left( \sum_{\tilde{W}'_\nu} A_{\nu, W'_\nu, \tilde{W}'_\nu} \hat{F}_\nu^{(W_\nu,\tilde{W}'_\nu)} - \sum_{\tilde{W}_\nu} A_{\nu, \tilde{W}_\nu, W_\nu} \hat{F}_\nu^{(\tilde{W}_\nu, W'_\nu)} \right),
\end{equation}
so that
\begin{align}
[\mcal{V}_{\hat{H}_c}]_{\lambda, W_\lambda, W'_\lambda; \nu, W_\nu, W'_\nu} & = \frac{i}{\hbar} \left( K_{\hat{H}}^{(\nu, W'_\nu,W'_\lambda)} \delta_{W_\lambda,W_\nu} - K_{\hat{H}}^{(\nu, W_\lambda, W_\nu)} \delta_{W'_\lambda,W'_\nu} \right) \delta_{\lambda, \nu}, \label{me1}\\
[\mcal{D}^{(\mathrm{loc})}_{\hat{\ell}}]_{\lambda, W_\lambda, W'_\lambda; \nu, W_\nu, W'_\nu} & = \gamma_{\mathrm{loc}} \bigg[ K_{\hat{\ell},\hat{\ell}}^{(\lambda, W_\lambda, W'_\lambda; \nu, W_\nu, W'_\nu)} \delta_{\lambda,\{\nu^{-+}\}} \label{me2} \\ & \qquad \qquad - \frac{1}{2} \left( K_{\hat{\ell}^\dagger \hat{\ell}}^{(\nu, W'_\nu, W'_\lambda)} \delta_{W_\lambda,W_\nu} + K_{\hat{\ell}^\dagger \hat{\ell}}^{(\nu, W_\lambda, W_\nu)} \delta_{W'_\lambda,W'_\nu}\right) \delta_{\lambda, \nu} \bigg], \nonumber \\
& \nonumber \\
[\mcal{D}^{(\mathrm{col})}_{\hat{L}}]_{\lambda, W_\lambda, W'_\lambda; \nu, W_\nu, W'_\nu} & = \gamma_c \bigg[ K_{\hat{L}}^{(\nu, W_\lambda, W_\nu)} K_{\hat{L}}^{(\nu, W'_\lambda, W'_\nu)\ast} \nonumber \\
& \qquad \quad - \frac{1}{2} \bigg( \sum_{\tilde{W}'_\nu} K_{\hat{L}}^{(\nu, \tilde{W}'_\nu, W'_\lambda)} K_{\hat{L}}^{(\nu, \tilde{W}'_\nu, W'_\nu)\ast} \bigg) \delta_{W_\lambda,W_\nu} \label{me3} \\
& \qquad \quad - \frac{1}{2} \bigg( \sum_{\tilde{W}_\nu} K_{\hat{L}}^{(\nu, \tilde{W}_\nu, W_\nu)} K_{\hat{L}}^{(\nu, \tilde{W}_\nu, W_\lambda)\ast} \bigg) \delta_{W'_\lambda,W'_\nu} \bigg] \delta_{\lambda,\nu}. \nonumber
\end{align}

If the local operators $\hat{\ell}$ and $\hat{L}$ are Hermitian, then so are the superoperators $\mcal{D}^{(\mathrm{loc})}_{\hat{\ell}}$ and $\mcal{D}^{(\mathrm{col})}_{\hat{L}}$~\cite{footnotea3}, i.e.,
\begin{equation}
\begin{aligned}
\relax[\mcal{D}^{(\mathrm{loc})}_{\hat{\ell}}]_{\nu, W_\nu, W'_\nu;\lambda, W_\lambda, W'_\lambda} & = [\mcal{D}^{(\mathrm{loc})}_{\hat{\ell}}]_{\lambda, W_\lambda, W'_\lambda; \nu, W_\nu, W'_\nu}^*, \\
[\mcal{D}^{(\mathrm{col})}_{\hat{L}}]_{\nu, W_\nu, W'_\nu;\lambda, W_\lambda, W'_\lambda} & = [\mcal{D}^{(\mathrm{col})}_{\hat{L}}]_{\lambda, W_\lambda, W'_\lambda; \nu, W_\nu, W'_\nu}^*.
\end{aligned}
\end{equation}

In addition, thanks to the symmetry relations (\ref{symrelCXY}), the following symmetry relation holds for \emph{any} local operators $\hat{\ell}$ and $\hat{L}$:
\begin{equation}
\begin{aligned}
\relax[\mcal{D}^{(\mathrm{loc})}_{\hat{\ell}}]_{\lambda, W'_\lambda, W_\lambda;\nu, W'_\nu, W_\nu} = [\mcal{D}^{(\mathrm{loc})}_{\hat{\ell}}]_{\lambda, W_\lambda, W'_\lambda; \nu, W_\nu, W'_\nu}^*, \\ [\mcal{D}^{(\mathrm{col})}_{\hat{L}}]_{\lambda, W'_\lambda, W_\lambda;\nu, W'_\nu, W_\nu} = [\mcal{D}^{(\mathrm{col})}_{\hat{L}}]_{\lambda, W_\lambda, W'_\lambda; \nu, W_\nu, W'_\nu}^*.
\end{aligned}
\end{equation}

\section{Application to qubit systems}

In this section, we exemplify our formalism for the qubit case and we show how Identity~1 of Ref.~\cite{Cha08} is directly recovered from the very general Eq.~(\ref{genIdentity1}) in the specific case $d = 2$. Qubit systems were handled in~\cite{Cha08} using a long inductive approach not extendable to multilevel systems.

For $d = 2$, the commutant basis operators $\hat{F}_\nu^{(W_\nu,W'_\nu)}$ are indexed with partitions $\nu \equiv (\nu_1,\nu_2) \in \mathcal{P}_2$ ($\nu_1 > 0, \nu_2 \geq 0$). In this case, the set $\{\nu^-\}$ is only composed of the valid partitions among the two partitions $\nu^{-_1} \equiv (\nu_1 - 1, \nu_2)$ and $\nu^{-_2} \equiv (\nu_1, \nu_2 - 1)$, so that $\{\nu^{-+}\}$ is in turn only composed of the valid partitions among the three partitions $\nu_a \equiv \nu$, $\nu_b \equiv \nu^{1 \rightarrow 2} = (\nu_1 - 1, \nu_2 + 1)$, and $\nu_c \equiv \nu^{2 \rightarrow 1} = (\nu_1 + 1, \nu_2 - 1)$~\cite{footnoteSnuUnud}. The cardinality of the sets $\mcal{W}_{\mu}^{(\pm j)}(W_\nu)$ and $\tilde{\mcal{W}}_{\mu}^{(\pm j)}(W_\nu)$ is at most $1$. Indeed, for $d = 2$ the content of the SWT boxes is either a 0 or a 1 and there is a unique SWT $W_\nu^{n_0}$ of shape $\nu$ with prescribed admissible content of $n_0$ boxes 0 and $n_1 = |\nu| - n_0$ boxes $1$ (the boxes $0$ have no other option than being located at the beginning of the first row of the SWT and the boxes 1 only on the rest). For all $j \in \{0, 1\}$, $\nu \in \mathcal{P}_2$, $\mu \in \{\nu^{\pm}\}$, and SWT $W_\nu^{n_0}$, we have $\mcal{W}_{\mu}^{(\pm j)}(W_\nu^{n_0}) = \tilde{\mcal{W}}_{\mu}^{(\pm j)}(W_\nu^{n_0}) = \{W_\mu^{n_0 \pm (1-j)}\}$  or $\emptyset$ (if $W_\mu^{n_0 \pm (1-j)}$ is not a valid SWT). As a result, $\forall \lambda, \mu, \nu : \{\lambda,\mu,\nu\} = 1$, $i,j \in \{0,1\}$, $\mcal{W}_\mu^{(-i)}(W_\lambda^{\tilde{n}_0}) \cap \mcal{W}_\mu^{(-j)}(W_\nu^{n_0}) = \{W_\mu^{\tilde{n}_0 + i - 1}\} \cap \{W_\mu^{n_0 + j - 1}\}$ and this set is not empty only if the two singletons coincide with a valid SWT, which at least requires $\tilde{n}_0 = n_0 + (j-i)$. In addition to the generic vanishing condition $\{\lambda, \mu, \nu\} = 0$, the single qubit operator $\hat{g}_\mu^{(\lambda,W_\lambda^{\tilde{n}_0};\nu,W_\nu^{n_0})}$ is necessarily zero if $|\tilde{n}_0 - n_0| > 1$.

Setting $n_0(q) = n_0 - q$, the only possibly nonzero $\hat{g}_\mu^{(\lambda,W_\lambda^{n_0(q)};\nu,W_\nu^{n_0})}$ operators are obtained for $q = 0, \pm 1$ (they can vanish within this condition for specific $\lambda$, $\mu$, $\nu$, $W_\lambda^{n_0(q)}$, and $W_\nu^{n_0}$). They are listed in Table~\ref{TableI}, along with their matrix elements, explicit expression, and a relevant trace property they fulfill. To this aim, we defined $\zeta_{k\tau}^{\nu, n_0} \equiv \langle W_{\nu^{-_\tau}}^{n_0 + k - 1}, k | W_\nu^{n_0} \rangle$ ($k = 0,1$, $\tau = 1, 2$), $\hat{s}_{+1} = |1\rangle \langle 0|$, $\hat{s}_{-1} = |0\rangle \langle 1|$, $\hat{s}_{0} = (|1\rangle \langle 1| - |0\rangle \langle 0|)/2$, and
\begin{equation}
A_q^{\nu,n_0} = \left \{
                                            \begin{array}{lcl}
                                               \sqrt{(\nu_1 - n_0 + 1)(n_0 - \nu_2)} & \mathrm{for} & q = 1 \\
                                               (\nu_1 + \nu_2 - 2 n_0)/2 &  & q = 0 \\
                                               \sqrt{(\nu_1 - n_0)(n_0 + 1 - \nu_2)} &  & q = -1
                                            \end{array}
                                            \right.,
\end{equation}

\begin{equation}
B_q^{\nu,n_0} = \left \{
                                            \begin{array}{lcl}
                                               \sqrt{(n_0 - \nu_2)(n_0 - \nu_2 - 1)} & \mathrm{for}  & q = 1 \\
                                               \sqrt{(\nu_1 - n_0)(n_0 - \nu_2)} &  & q = 0 \\
                                               -\sqrt{(\nu_1 - n_0 - 1)(\nu_1 - n_0)} &  & q = -1
                                            \end{array}
                                            \right.,
\end{equation}

\begin{equation}
D_q^{\nu,n_0} = \left \{
                                            \begin{array}{lcl}
                                               -\sqrt{(\nu_1 - n_0 + 1)(\nu_1 - n_0 + 2)} & \mathrm{for}  & q = 1 \\
                                               \sqrt{(\nu_1 - n_0 + 1)(n_0 - \nu_2 + 1)} &  & q = 0 \\
                                               \sqrt{(n_0 - \nu_2 + 1)(n_0 - \nu_2 + 2)} &  & q = -1
                                            \end{array}
                                            \right..
\end{equation}
The Kronecker delta in the third column of Table~\ref{TableI} accounts for the necessary condition $n_0(q) = n_0 + j - i$ for the matrix elements to be nonzero. The operator expressions in the fourth and fifth columns directly follow from the explicit expressions of the CGC's $\zeta_{k\tau}^{\nu, n_0}$ for all $\nu = (\nu_1,\nu_2) \in \mathcal{P}_2$, i.e. (see Eq.~(\ref{CGCd}) with $d = 2$), 
\begin{equation}
\begin{aligned}
    \zeta_{01}^{\nu,n_0} & = \sqrt{\frac{n_0 - \nu_2}{\Delta \nu}}, \quad \zeta_{11}^{\nu,n_0} = \sqrt{\frac{\nu_1 - n_0}{\Delta \nu}}, \\ \zeta_{02}^{\nu,n_0} & = - \sqrt{\frac{\nu_1 + 1 - n_0}{\Delta \nu + 2}}, \quad \zeta_{12}^{\nu,n_0} = \sqrt{\frac{n_0 - \nu_2 + 1}{\Delta \nu + 2}},
\end{aligned}
\end{equation}
where $\Delta \nu \equiv \nu_1 - \nu_2$. The trace property in the sixth column merely stems from the elementary relations $\mathrm{Tr}[\hat{s}_q] = 0$, $\mathrm{Tr}[\hat{s}_{\pm 1}^\dagger \hat{s}_q] = \delta_{q,\pm 1}$, and $\mathrm{Tr}[\hat{s}_0^\dagger \hat{s}_q] = \delta_{q,0}/2$, $\forall q = 0,\pm 1$.



\begin{table}
\renewcommand{\arraystretch}{1.5}
\begin{tabular}{c|c|c|c|c|c}
$\lambda$ & $\mu$ & $[g_\mu^{(\lambda,W_\lambda^{n_0(q)};\nu,W_\nu^{n_0})}]_{i,j}$ & $\hat{g}_\mu^{(\lambda,W_\lambda^{n_0(\pm 1)};\nu,W_\nu^{n_0})}$ & $\hat{g}_\mu^{(\lambda,W_\lambda^{n_0(0)};\nu,W_\nu^{n_0})}$ & $\mathrm{Tr}[\hat{g}_\mu^{(\lambda,W_\lambda^{n_0(q')};\nu,W_\nu^{n_0})\dagger}\hat{s}_q]$ \\
\hline
$\nu$ & $\nu^{-_1}$ & $\zeta_{i 1}^{\nu,n_0(q)} \zeta_{j 1}^{\nu,n_0} \delta_{q,i-j}$ & $\frac{A_{\pm 1}^{\nu, n_0}}{\Delta \nu} \hat{s}_{\pm 1}$ & $\sum_{k= 
 0}^1 [\zeta_{k1}^{\nu,n_0}]^2|k\rangle \langle k|$ & $\frac{A_q^{\nu,n_0}}{\Delta \nu} \delta_{q,q'}$ \\
& $\nu^{-_2}$ & $\zeta_{i 2}^{\nu,n_0(q)} \zeta_{j 2}^{\nu,n_0} \delta_{q,i-j}$ & $-\frac{A_{\pm 1}^{\nu, n_0}}{\Delta \nu+2} \hat{s}_{\pm 1}$ & $\sum_{k= 
 0}^1 [\zeta_{k2}^{\nu,n_0}]^2|k\rangle \langle k|$ & $-\frac{A_q^{\nu,n_0}}{\Delta \nu + 2} \delta_{q,q'}$ \\
$\nu_b$ & $\nu^{-_1} = \nu_b^{-_2}$ & $\zeta_{i 2}^{\nu_b,n_0(q)} \zeta_{j 1}^{\nu,n_0} \delta_{q,i-j}$ & $\frac{B_{\pm 1}^{\nu, n_0}}{\Delta \nu} \hat{s}_{\pm 1}$ & $2 \frac{B_0^{\nu, n_0}}{\Delta \nu} \hat{s}_0$ & $\frac{B_q^{\nu,n_0}}{\Delta \nu} \delta_{q,q'}$ \\
$\nu_c$ & $\nu^{-_2} = \nu_c^{-_1}$ & $\zeta_{i 1}^{\nu_c,n_0(q)} \zeta_{j 2}^{\nu,n_0} \delta_{q,i-j}$ & $\frac{D_{\pm 1}^{\nu, n_0}}{\Delta \nu + 2} \hat{s}_{\pm 1}$ & $2 \frac{D_0^{\nu, n_0}}{\Delta \nu + 2} \hat{s}_0$ & $\frac{D_q^{\nu,n_0}}{\Delta \nu + 2} \delta_{q,q'}$
\end{tabular}
\caption{Only possibly nonzero operators $\hat{g}_\mu^{(\lambda,W_\lambda^{n_0(q)};\nu,W_\nu^{n_0})}$ ($q = 0, \pm 1$) for a given $\nu \equiv (\nu_1,\nu_2) \in \mathcal{P}_2$ and $W_\nu^{n_0} \in \mcal{W}_\nu$.\label{TableI}}
\end{table}

Finally, we have for all $\nu \equiv (\nu_1, \nu_2) \vdash (N,2)$~\cite{footnotefnu}
\begin{equation}
    f^\nu = \frac{\Delta \nu + 1}{\nu_1 + 1} \binom{N}{\nu_2},
\end{equation}
so that
\begin{equation}
    r_{\nu}^{\nu^{-_1}} = \frac{\Delta \nu}{\Delta \nu + 1}(\nu_1 + 1), \quad r_{\nu}^{\nu^{-_2}} = \frac{\Delta \nu + 2}{\Delta \nu + 1}\nu_2
\end{equation}
and
\begin{equation}
\frac{f^\nu}{f^{\nu_b}} = \frac{(\nu_2 + 1)(\Delta \nu + 1)}{(\nu_1 + 1)(\Delta \nu - 1)}, \quad \frac{f^\nu}{f^{\nu_c}} = \frac{(\nu_1 + 2)(\Delta \nu + 1)}{\nu_2(\Delta \nu + 3)}.
\end{equation}

As a result, for $\hat{X} = \hat{s}_q$ and $\hat{Y} = \hat{s}_r$ ($q,r = 0, \pm 1$) the general Identity~(\ref{genIdentity1}) straightforwardly simplifies to
\begin{equation}
\begin{aligned}
\sum_{n = 1}^N \hat{s}_q^{(n)} \overline{|\nu, W_\nu^{n_0}\rangle \langle \nu, W_\nu^{n'_0}|} \hat{s}_r^{(n)\dagger} = {} & \frac{\nu_1 + \nu_2 + 2}{\Delta \nu (\Delta \nu + 2)} A_q^{\nu,n_0} A_r^{\nu,n'_0} \overline{|\nu, W_\nu^{n_0(q)}\rangle \langle \nu, W_\nu^{n'_0(r)}|} \\
 & \hspace{-3cm} + \frac{\nu_1 + 1}{\Delta \nu (\Delta \nu + 1)} B_q^{\nu,n_0} B_r^{\nu,n'_0} \sqrt{\frac{f^\nu}{f^{\nu_b}}} \overline{|\nu_b, W_{\nu_b}^{n_0(q)}\rangle \langle \nu_b, W_{\nu_b}^{n'_0(r)}|} \label{idqubitFinal}
 \\ 
& \hspace{-3cm} + \frac{\nu_2}{(\Delta \nu + 1)(\Delta \nu + 2)} D_q^{\nu,n_0} D_r^{\nu,n'_0} \sqrt{\frac{f^\nu}{f^{\nu_c}}} \overline{|\nu_c, W_{\nu_c}^{n_0(q)}\rangle \langle \nu_c, W_{\nu_c}^{n'_0(r)}|}.
\end{aligned}
\end{equation}

For $d = 2$, the Schur basis states $|\nu, T_\nu, W_\nu\rangle$ are nothing but the standard Clebsch-Gordan basis states $|J,M,i\rangle$, according to the correspondance $J = \Delta \nu/2$, $M = N/2 - n_0$ (equivalently $\nu_1 = N/2 + J$, $\nu_2 = N/2 - J$, and $n_0 = N/2 - M$), and where $i$ is indexed by the distinct SYT's $T_\nu$, hence from $1$ to $f^\nu = \binom{N}{N/2 - J}(2J + 1)/(J + 1 + N/2) \equiv d_N^J$. Equation (\ref{idqubitFinal}) is just Identity 1 of Ref.~\cite{Cha08} expressed in the Schur-Weyl duality language. For $(\nu, W_\nu^{n_0}) \equiv (J,M)$, $(\nu, W_{\nu}^{n_0(q)}) = (J,M_q)$, $(\nu_b, W_{\nu_b}^{n_0(q)}) = (J - 1,M_q)$ and $(\nu_c, W_{\nu_c}^{n_0(q)}) = (J + 1,M_q)$, with $M_q = M + q$.

\section{Conclusion}
\label{ConclSection}

In this paper, we established the general theoretical framework that allows for an exact description of the open system dynamics of permutationally invariant states in arbitrary $N$-qudit systems ($d \geq 2$) when constrained within the commutant $\mathcal{L}_{S_N}(\mcal{H})$ (the subspace of permutationally invariant operators in the global Liouville space of operators acting in the system Hilbert space). Thanks to Schur-Weyl duality powerful results, we identified a natural orthonormal basis of operators in the commutant onto which the master equation can be projected and provided the exact expansion coefficients in the most general case (arbitrary dynamics, $N$ and $d$). While we here specifically focused on general time-local Markovian or non-Markovian master equations, our formalism can also be applied to more general master equations that would make use of any permutationally invariant linear maps of the form of Eq.~(\ref{KXYdef}) for which we identified the exact matrix elements in the natural orthonormal basis of  the commutant. The formalism does not require one to compute the Schur transform and allows to remain completely restricted within the commutant subspace, whose dimension only scales polynomially with the number $N$ of qudits instead of exponentially, as is the case for the whole Liouville space of operators. We introduced the concept of $3\nu$-symbol matrix that proves to be particularly useful in this context. We finally showed how our theoretical framework particularizes for qubit systems ($d = 2$) and how previously known results are recovered for this specific case. Being exact and focused on the commutant subspace, our formalism should make it possible to simulate the dynamics of open many-body quantum systems with a large number of constituents more efficiently than is possible today.

The versatility of our formalism opens up a broad spectrum of potential applications across various domains, establishing it as a flexible tool for exploring novel quantum phenomena, some of which may be specific to ensemble of qudits. It enables the exploration of the interplay between quantum driving, collective dissipation, and individual dissipation, which could provide insights into the essential factors contributing to the emergence of dissipative time crystals~\cite{San20, Pas22}. It is particularly suited to examining the impact of individual losses, ubiquitous in experimental setups, on collective systems of qudits~\cite{Lew19}. Models based on two-level systems can find experimental realisation in configurations that involve individual quantum systems with more than two levels, as seen in cases such as trapped ions. Consequently, there is significant interest in evaluating the losses from these additional levels. Our formalism is distinct in its group-theoretic analytical treatment of dissipative dynamics and in the exact results it yields, which could be particularly relevant in the context of quantum thermodynamics~\cite{Yad2023}. It complements numerous other methods relying e.g.\ on stochastic quantum trajectories or different ansatzes for the density operator entering the Lindblad-like master equation, such as in cluster mean-field approaches~\cite{Jin16}, variational neural-network ansatz~\cite{Nag19,Har19,Vic19} or tensor-network techniques. Furthermore, our formalism finds relevance in studying the impact of noise on quantum circuits utilizing qudits, which is particularly useful for advancing progress in noisy intermediate scale quantum devices (NISQ)~\cite{Li2022} or in quantum metrology by analyzing spin squeezing in ensembles of spin-$s$ particles and the consequences of imperfect preparation of the initial state, as was done for two-level systems~\cite{Wes02}. It also provides a powerful tool for exploring the strong coupling of cavity modes to collective excitations in ensembles of quantum emitters, accounting for the discrete multilevel spectrum of molecular systems beyond the Tavis-Cummings model. This consideration is crucial for understanding the complex interplay of molecular electronic and nuclear degrees of freedom, particularly in the context of light-matter hybrid states known as polaritons~\cite{Cam22,Pre23}. Additionally, our formalism is well-suited for investigating measurement-induced phase transitions in systems with more than two-level quantum components, which arises from the competition between quantum measurements and coherent dynamics~\cite{Sie22}. 

\section*{Data availability statement}
 All data that support the findings of this study are included within the article (and any supplementary files).

\ack
This project (EOS 40007526) has received funding from the FWO and F.R.S.-FNRS under the Excellence of Science (EOS) programme. T.B. also acknowledges financial support through IISN convention 4.4512.08.

\appendix

\section{Dimension of the commutant $\mathcal{L}_{S_N}(\mcal{H})$}
\label{dimCommutantApp}
The commutant $\mathcal{L}_{S_N}(\mcal{H})$ coincides with the symmetric subspace of the Liouville space $\mathcal{L}(\mcal{H})$~\cite{footnote3b} and its dimension thus reads $f^{(N)}(d^2) = \binom{N + d^2 - 1}{N}$. On the other hand, as a result of the Schur-Weyl duality, this dimension also identifies to $\sum_{\nu \vdash(N,d)}f^\nu(d)^2$, with $f^\nu(d)$ the dimension of the $\mcal{U}^\nu(d)$ irrep of the general linear group $GL(d)$ (and of the unitary group $U(d)$). Both expressions must of course coincide and this can be easily seen as follows.

We first observe that $\sum_{\nu \vdash(N,d)}f^\nu(d)^2$ can also be written $\sum_{\nu \vdash(N,d)}s_\nu(1, \ldots, 1)^2$, where $(1, \ldots, 1)$ is a $d$-uple and $s_\nu(x_1, \ldots, x_d)$ is the Schur's polynomial in the $d$ variables $x_1, \ldots, x_d$ associated to partition $\nu$ (this is an homogeneous symmetric polynomial of degree $N$ in the variables $x_1, \ldots, x_d$ - see~\cite{footnotefnud}). For all $\mathbf{x} \equiv (x_1, \ldots, x_d)$ and $\mathbf{y} \equiv (y_1, \ldots, y_d)$, Cauchy's identity states that
\begin{equation}
    \sum_{N = 0}^\infty \sum_{\nu \vdash (N,d)} s_\nu(\mathbf{x})s_\nu(\mathbf{y}) = \prod_{i,j = 1}^d (1 - x_i y_j)^{-1},
\end{equation}
which for $\mathbf{x} = \mathbf{y} = (t, \ldots, t)$ particularizes to
\begin{equation}
    \sum_{N = 0}^\infty \sum_{\nu \vdash (N,d)} s_\nu(t, \ldots, t)^2 = (1 - t^2)^{-d^2}.
    \label{Cid}
\end{equation}

On the one hand, $s_\nu(t, \ldots, t)$ is a multiple of $t^N$ for all $\nu \vdash(N,d)$, so that
\begin{equation}
\sum_{\nu \vdash (N,d)} s_\nu(t, \ldots, t)^2 = \alpha_{N,d} t^{2 N}, \quad \forall t,
\label{snutd}
\end{equation}
with $\alpha_{N,d}$ a real constant that depends on $N$ and $d$. On the other hand, $(1 - t^2)^{-d^2}$ admits for $-1 < t < 1$ the series expansion $\sum_{N = 0}^\infty \binom{N + d^2 - 1}{N} t^{2 N}$. As a result, Eq.~(\ref{Cid}) can be written for $-1 < t < 1$
\begin{equation}
    \sum_{N = 0}^{\infty} \alpha_{N,d} t^{2 N} = \sum_{N = 0}^\infty \binom{N + d^2 - 1}{N} t^{2 N}
\end{equation}
and $\alpha_{N,d}$ must identify to $\binom{N + d^2 - 1}{N}, \forall N,d$. It then follows from Eq.~(\ref{snutd}) that
\begin{equation}
\sum_{\nu \vdash (N,d)} s_\nu(1, \ldots, 1)^2 = \binom{N + d^2 - 1}{N}.
\end{equation}

\section{Clebsch-Gordan coefficients of tensor products of unitary group irreps in the context of qudit systems}
\label{CGApp}

In this Appendix, the CGC's of tensor products of unitary group irreps are explicitly detailed in the specific contexts of qudit systems with $d$ levels denoted by $|0\rangle, \ldots, |d-1\rangle$ (to comply with the standard qubit level notation $|0\rangle$ and $|1\rangle$ if $d = 2$).

The tensor product of two (unitary) irreps $\mcal{U}^\mu(d)$ and $\mcal{U}^\nu(d)$ of the unitary group $U(d)$ ($\mu, \nu \in \mathcal{P}_d$) decomposes into a direct sum of irreducible components according to~\cite{Fulton}
\begin{equation}
    \mcal{U}^\mu(d) \otimes \mcal{U}^\nu(d) = \bigoplus_{\lambda \in \mathcal{P}_d} \bigoplus_{r = 1}^{c^\lambda_{\mu \nu}} \mcal{U}^{\lambda}(d)_r,
    \label{tensorproductirreps}
\end{equation}
with $c^\lambda_{\mu \nu}$ the so-called Littlewood-Richardson coefficients and $\mcal{U}^{\lambda}(d)_r$ ($\lambda \in \mathcal{P}_d, r = 1, \ldots, c^\lambda_{\mu \nu}$) equivalent $\mcal{U}^{\lambda}(d)$-irreps of $U(d)$. The index $r$ is omitted if it only takes value 1. The Clebsch-Gordan coefficients (CGC's) of the tensor product $\mcal{U}^\mu(d) \otimes \mcal{U}^\nu(d)$ for the Gel'fand-Tsetlin (GT) basis are the expansion coefficients $\langle W_\mu, W_\nu | W_\lambda \rangle_r$ of the GT-basis vectors $|W_{\lambda}\rangle_r$ of $\mcal{U}^{\lambda}(d)_r$ in the basis formed by the tensor products of the GT-basis vectors $|W_\mu\rangle$ of $\mcal{U}^\mu(d)$ with the GT-basis vectors $|W_\nu\rangle$ of $\mcal{U}^\nu(d)$:
\begin{equation}
    |W_{\lambda}\rangle_r = \sum_{W_\mu, W_\nu} \langle W_\mu, W_\nu | W_\lambda \rangle_r |W_\mu\rangle \otimes |W_\nu\rangle, \qquad \forall \lambda \in \mathcal{P}_d, W_\lambda \in \mcal{W}_\lambda, r.
\end{equation}
The decomposition (\ref{tensorproductirreps}) is not unique as soon as any of the Littlewood-Richardson coefficients $c^\lambda_{\mu \nu}$ exceeds 1, in which case the CGC's are neither univocally defined.

For $\nu = (1)$, Eq.~(\ref{tensorproductirreps}) specifically reads
\begin{equation}
    \mcal{U}^\mu(d) \otimes \mcal{U}^{(1)}(d) = \bigoplus_{\lambda \in \{\mu^+\}} \mcal{U}^\lambda(d)
    \label{tensorproductirreps2}
\end{equation}
and the CGC's are univocally determined. The $d$-dimensional representation $\mcal{U}^{(1)}(d)$ can always be chosen to have each unitary matrix $U \in U(d)$ be represented by a unitary operator having its representation matrix in the orthonormal basis $|0\rangle, \ldots, |d - 1\rangle$ of $\mcal{U}^{(1)}(d)$ given by $U$ (standard representation of $U(d)$ on the single qudit state space $\mcal{H}_d \cong \mcal{U}^{(1)}(d)$). If we view the $U(d)$ subgroups $U(k)$ ($1 \leq k < d$) as the groups of $d \times d$ unitary matrices $U_k \oplus (1)^{\oplus(d - k)}$ ($U_k$ $k \times k$ unitary subblocks) that leave fixed the basis vectors $|k\rangle, \ldots, |d-1\rangle$ in the representation $\mcal{U}^{(1)}(d)$, the GT-basis vectors $|W_{(1)}\rangle$ of $\mcal{U}^{(1)}(d)$ merely identify to the orthonormal basis vectors $|j\rangle$ ($j = 0, \ldots, d-1$) and the CGC's $\langle W_\mu, W_{(1)} | W_\lambda \rangle$ can be accordingly denoted by $\langle W_\mu, j | W_\lambda \rangle$:
\begin{equation}
    |W_\lambda\rangle = \sum_{j = 0}^{d-1} \sum_{W_\mu \in \mcal{W}_\mu} \langle W_\mu, j | W_\lambda \rangle |W_\mu\rangle \otimes |j\rangle, \qquad \forall \lambda \in \{\mu^+\}, W_\lambda \in \mcal{W}_\lambda.
    \label{couplingnupnu}
\end{equation}

The CGC's $\langle W_\mu, j | W_\lambda \rangle$ are expressed in terms of the Gel'fand-Tsetlin (GT) patterns of the SWT's $W_\mu$~\cite{GelfandTsetlin, Vilenkin}. For all $\nu \in \mathcal{P}_d$, the GT pattern $G(W_\nu)$ of a SWT $W_\nu$ is the $d$-row tableau listing in a one row - one partition format all partitions $\nu(k)$ ($k = 1, \ldots, d$) of shapes of $W_\nu$ where only boxes $0$, \ldots, $k-1$ are kept. The partitions are listed in the reversed order $k = d, \ldots, 1$. Each $\nu(k)$ is a partition of at most $k$ parts, denoted standardly by the numbers $m_{i,k}$, $i = 1, \ldots, k$, set to $0$ for all $i$ that exceed the length of partition $\nu(k)$. The form of a GT pattern only depends on $d$ and is an inverse triangular pattern of $d$ numbers on the first line, $d-1$ on the second, \ldots, and finally 1 on the last $d$th line, whatever the partition $\nu \in \mathcal{P}_d$ and the SWT $W_\nu$:
\begin{equation}
    G(W_\nu) \equiv \left(
    \begin{array}{ccccccccc}
        m_{1,d} & & \!\!\!\!\!\!m_{2,d} & & \!\!\!\!\!\!\ldots & & \!\!\!\!\!\!m_{d-1,d} & & \!\!\!\!\!\!m_{d,d} \\
            & \!\!\!\!\!\!m_{1, d-1} & & \!\!\!\!\!\!\ldots & & \!\!\!\!\!\!\ldots & & \!\!\!\!\!\!m_{d-1,d-1} & \\
            & & & & \!\!\!\!\!\!\ldots & & & & \\
            & & & \!\!\!\!\!\!m_{1,2} & & \!\!\!\!\!\!m_{2,2} & & & \\
            & & & & \!\!\!\!\!\!m_{1,1} & & & &
    \end{array}
    \right).
    \label{Gelfand}
\end{equation}
The numbers $m_{i,d}$ ($i = 1, \ldots, d$) on the top line merely identify to the parts $\nu_i$ of the partition $\nu$ of the SWT $W_\nu$, possibly completed with zeroes if the partition length is lower than $d$. The GT pattern $G(W_\nu)$ is either denoted accordingly by $\binom{\nu}{m}$, or also merely by $(m)$. For all $k = 1, \ldots, d$, the number $\alpha_k = |m_k| - |m_{k-1}|$, with $|m_k| \equiv \sum_{i = 1}^k m_{i,k}$ and $m_0 \equiv 0$, yields the number $n_{k-1}$ of boxes $(k-1)$ in the SWT $W_\nu$. In particular $n_0 = m_{1,1}$. The numbers $m_{i,k}$ satisfy the betweenness condition: $m_{i,k-1} \in [m_{i+1,k},m_{i,k}]$, $\forall k = 2, \ldots, d; i = 1, \ldots, k-1$. Any triangular pattern satisfying the betweenness condition is by definition a GT pattern. For all partition $\nu$, there is a one to one correspondence between the set of all GT patterns $\binom{\nu}{m}$ and the set of all SWT's $W_\nu$, so that the knowledge of all numbers $m_{i,k}$ uniquely identify a SWT. In particular, for $d = 2$, the GT pattern for any SWT $W_{\nu \equiv (\nu_1,\nu_2)}$ merely reads
\begin{equation}
    G(W_\nu) = \left(
    \begin{array}{ccc}
         \nu_1 & & \nu_2 \\
          & n_0 &
    \end{array}
    \right).
\end{equation}

For all $i,\tau = 1, \ldots, d$, a triangular shift pattern $(i, \tau)$, $\Delta_i(\tau, \tau_{d-1}, \ldots, \tau_i)$, is a pattern of $d$ rows containing only 0's and 1's according to
\begin{equation}
    \Delta(\tau, \tau_{d-1}, \ldots, \tau_i) =
    \left(
    \begin{array}{c}
        e_{\tau} \\ e_{\tau_{d - 1}} \\ \vdots \\ e_{\tau_i} \\ (0)_{i - 1}
    \end{array}
    \right),
\end{equation}
where, $\forall k = i, \ldots, d-1$, $1 \leq \tau_k \leq k$, $e_{\tau_k}$ is a unit row vector of length $k$ with 1 at position $\tau_k$ and 0 elsewhere, and $(0)_{i-1}$ denotes a triangular array of zeroes with $i - 1$ rows. The set of all triangular shift patterns $(i, \tau)$ for all possible values of $\tau_{d-1}, \ldots, \tau_i$ is denoted by $\Delta_i(\tau)$. It is composed of $(d-1)!/(i-1)!$ elements. For $d = 2$ or also $i = d$, this is therefore a singleton.

We define the difference of two GT patterns $(m)$ and $(m')$ by the triangular pattern of elements $m_{i,k} - m'_{i,k}$. If $W_\mu$ and $W_\nu$ are two SWT's such that
\begin{equation}
G(W_\nu) - G(W_\mu) \in \Delta_i(\tau),
\label{cond}
\end{equation}
then $\nu = \mu^{+_\tau}$ (equivalently $\mu = \nu^{-_\tau}$)~\cite{footnotenumtau} and both $W_\mu$ and $W_\nu$ have globally identical content up to one more box $(i-1)$ in $W_\nu$. Indeed, setting $G(W_\mu) \equiv (m)$ and $G(W_\nu) \equiv (m')$, we then have $|m'_k| = |m_k| + 1, \forall k \geq i$ and $|m'_k| = |m_k|, \forall k < i$, so that $\alpha'_k = \alpha_k$, $\forall k \neq i$ and $\alpha'_i = \alpha_i + 1$. For all $\mu,\nu \in \mathcal{P}_d : \nu \in \{\mu^+\}$ ($\Leftrightarrow \mu \in \{\nu^-\}$), $j = 0, \ldots, d-1$, and SWT $W_\mu$, we denote accordingly by $\mcal{W}_\nu^{(+j)}(W_\mu)$ the set of all SWT's $W_\nu$ of shape $\nu$ that fullfill condition (\ref{cond}) with $i = j + 1$ and $\tau = \tau_{\nu/\mu}$~\cite{footnotetaunumu}. This is a subset of the set $\tilde{\mcal{W}}_\nu^{(+j)}(W_\mu)$ of \emph{all} SWT's of shape $\nu$ and same content as $W_\mu$ plus one box $j$. Similarly, for all SWT $W_\nu$, we denote by $\mcal{W}_\mu^{(-j)}(W_\nu)$ the set of all SWT's $W_\mu$ of shape $\mu$ that fullfill condition (\ref{cond}) with $i = j + 1$ and $\tau = \tau_{\mu/\nu}$. This is a subset of the set $\tilde{\mcal{W}}_\mu^{(-j)}(W_\nu)$ of \emph{all} SWT's of shape $\mu$ and same content as $W_\nu$ minus one box $j$. The subset is empty if $W_\nu$ does not contain any box $j$. We have
\begin{equation}
    W_\nu \in \mcal{W}_\nu^{(+j)}(W_\mu) \quad \Leftrightarrow \quad W_{\mu} \in \mcal{W}_\mu^{(-j)}(W_\nu).
    \label{equivalenceDelta}
\end{equation}
The cardinality of the subsets $\mcal{W}_\nu^{(+ j)}(W_\mu)$ and $\mcal{W}_\mu^{(-j)}(W_\nu)$ is identical and at most $\Delta_{j+1}(\tau_{\nu/\mu})$'s one, i.e., $(d-1)!/j!$ ($1$ for $d = 2$ or $j = d - 1$).

The CGC's $\langle W_\mu, j | W_\lambda \rangle$ in Eq.~(\ref{couplingnupnu}) are zero if $G(W_\lambda) - G(W_\mu) \notin \Delta_{j+1}(\tau_{\lambda/\mu})$, and otherwise
\begin{equation}
    \langle W_\mu, j | W_\lambda \rangle = \left[ \frac{\displaystyle \prod_{k = 1}^j (p_{\tau_{j+1},j+1} - p_{k,j})}{\displaystyle \prod_{\stackrel{k = 1}{k \neq \tau_{j+1}}}^{j+1} (p_{\tau_{j+1},j+1} - p_{k,j+1})} \right]^{1/2} \prod_{l = j+2}^d A_{\tau_{l-1},\tau_l},
    \label{CGCd}
\end{equation}
where we set $G(W_\lambda) - G(W_\mu) \equiv \Delta(\tau_d, \tau_{d-1}, \ldots, \tau_{j+1})$ (with $\tau_d = \tau_{\lambda/\mu}$), $G(W_\mu) \equiv (m)$, $p_{i,k} = m_{i,k} + k - i$, and where
\begin{equation}
    A_{\tau_{l-1},\tau_l} = \mathrm{sgn}(\tau_{l-1} - \tau_l) \left[ \prod_{\stackrel{k = 1}{k \neq \tau_l}}^{l} \frac{p_{\tau_{l-1},l-1} - p_{k,l} + 1}{p_{\tau_l,l} - p_{k,l}} \prod_{\stackrel{k = 1}{k \neq \tau_{l-1}}}^{l-1} \frac{p_{\tau_l,l} - p_{k,l-1}}{p_{\tau_{l-1,l-1} - p_{k,l-1} + 1}} \right]^{1/2},
\end{equation}
with $\mathrm{sgn}$ the sign function with $\mathrm{sgn}(0) = 1$. The CGC's are real and obey for all $\mu \in \{\lambda^-\} \cap \{\nu^-\}$ the orthogonality relation~\cite{Vilenkin}
\begin{equation}
    \label{orthogonality}
    \sum_{j=0}^{d - 1}\sum_{W_\mu \in \mcal{W}_\mu} \langle W_\mu, j |W_\lambda\rangle \langle W_\mu, j |W_\nu\rangle = \delta_{\lambda,\nu} \delta_{W_\lambda,W_\nu}.
\end{equation}

Equation~(\ref{couplingnupnu}) can be accordingly refined to
\begin{equation}
    |W_\lambda\rangle = \sum_{j = 0}^{d-1} \sum_{W_\mu \in \mcal{W}_\mu^{(-j)}(W_\lambda)} \langle W_\mu, j | W_\lambda \rangle |W_\mu\rangle \otimes |j\rangle, \quad \forall \lambda \in \{\mu^+\}, W_\lambda \in \mcal{W}_\lambda.
    \label{couplingnupnu2}
\end{equation}

\section{Proof of general Identity [Eq.~(\ref{genIdentity1})]}
\label{generalIdentityApp}
For any local operators $\hat{X}$ and $\hat{Y}$, $\mcal{K}_{\hat{X},\hat{Y}}[\hat{F}_\nu^{(W_\nu,W'_\nu)}]$ is a PI operator and can be expanded according to
\begin{equation}
\mcal{K}_{\hat{X},\hat{Y}}[\hat{F}_\nu^{(W_\nu,W'_\nu)}] = \sum_{\lambda \vdash (N,d)} \sum_{W_\lambda, W'_\lambda \in \mcal{W}_\lambda}  \mathrm{Tr}(\hat{F}_\lambda^{(W_\lambda,W'_\lambda)\dagger} \mcal{K}_{\hat{X},\hat{Y}}[\hat{F}_\nu^{(W_\nu,W'_\nu)}]) \hat{F}_\lambda^{(W_\lambda,W'_\lambda)}.
\end{equation}

We first observe that for any PI operator $\hat{A}_\mathrm{PI}$ and any operator $\hat{B}$ we have $\mathrm{Tr}(\hat{A}_\mathrm{PI}^\dagger \mcal{P}_\sigma[\hat{B}]) = \mathrm{Tr}(\hat{A}_\mathrm{PI}^\dagger \hat{B}), \forall \sigma \in S_N$~\cite{footnotea1}. As a result and since $\mcal{P}_\sigma[\hat{X}^{(n)} \hat{A}_\mathrm{PI} \hat{Y}^{(n)\dagger}] = \hat{X}^{(\sigma(n))} \hat{A}_\mathrm{PI} \hat{Y}^{(\sigma(n))\dagger}$, $\forall n, \sigma$, $\mathrm{Tr}(\hat{F}_\lambda^{(W_\lambda,W'_\lambda)\dagger} \hat{X}^{(n)} \hat{A}_\mathrm{PI} \hat{Y}^{(n)\dagger})$ is independent of $n$ and we have
\begin{equation}
\begin{aligned}
\mathrm{Tr}(\hat{F}_\lambda^{(W_\lambda,W'_\lambda)\dagger} \mcal{K}_{\hat{X},\hat{Y}}[\hat{F}_\nu^{(W_\nu,W'_\nu)}])
& \\
& \hspace{-4cm} = N \mathrm{Tr}(\hat{F}_\lambda^{(W_\lambda,W'_\lambda)\dagger} \hat{X}^{(N)} \hat{F}_\nu^{(W_\nu,W'_\nu)} \hat{Y}^{(N)\dagger}) \\
& \hspace{-4cm} = \frac{N}{\sqrt{f^\lambda f^\nu}} \sum_{T_\lambda, T_\nu} \langle \lambda, T_\lambda, W_\lambda|\hat{X}^{(N)}|\nu, T_\nu, W_\nu\rangle \langle \lambda, T_\lambda, W'_\lambda|\hat{Y}^{(N)}|\nu, T_\nu, W'_\nu\rangle^*.
\label{gen}
\end{aligned}
\end{equation}

The $N$-qudit system state space can be structured along $\mcal{H} = (\otimes_i^{N-1} \mcal{H}_i) \otimes \mcal{H}_N$. This global structure can further be refined thanks to the specific properties of the Schur basis states. According to Pieri's rule for the unitary group irreducible representations [Eq.~(\ref{tensorproductirreps2})] and considering the chain of irreps of $S_N, \ldots, S_1$ each Schur-Weyl basis state belongs to (here, $S_k$ ($1 \leq k < N$) denotes the $S_N$ subgroup that fixes each $j \in \{k+1, \ldots, N\}$ and only permutes $\{1,\ldots,k\}$), we get that, for all $\nu \vdash (N,d)$ and $T_\nu \in \mcal{T}_\nu$, $\mcal{H}_\nu(T_\nu) \simeq \mcal{U}^\nu(d)$ is an irreducible component of $\mcal{H}_{\nu(N-1)}(T_{\nu(N-1)}) \otimes \mcal{H}_N \simeq \mcal{U}^{\nu(N-1)}(d) \otimes \mcal{U}^{(1)}(d)$, with $\nu(N-1)$ the shape of $T_\nu$ without box $N$ and $T_{\nu(N-1)}$ the SYT $T_\nu$ without box $N$. This implies
\renewcommand{\arraystretch}{1.3}
\begin{equation}
    |\nu, T_\nu, W_\nu\rangle = \sum_{j=0}^{d-1} \sum_{\mbox{\scriptsize $\begin{array}{c} W_{\nu(N-1)} \in \\ \mcal{W}_{\nu(N-1)}^{(-j)}(W_\nu) \end{array}$}} \langle W_{\nu(N-1)}, j | W_\nu \rangle |\nu(N-1), T_{\nu(N-1)}, W_{\nu(N-1)}\rangle \otimes |j\rangle_N,
    \label{id}
\end{equation}
\renewcommand{\arraystretch}{1}
where the coefficients $\langle W_{\nu(N-1)}, j | W_\nu \rangle$ are the (real) Clebsch-Gordan coefficients (CGC's) of the tensor product $\mcal{U}^{\nu(N-1)}(d) \otimes \mcal{U}^{(1)}(d)$ for the Gel'fand-Tsetlin bases. These coefficients are zero iff $W_{\nu(N-1)} \notin \mcal{W}_{\nu(N-1)}^{(-j)}(W_\nu)$, so that the sum in Eq.~(\ref{id}) can be formally extended to the whole set $\mcal{W}_{\nu(N-1)}$. Proceeding similarly for expanding the state $|\lambda, T_\lambda, W_\lambda\rangle$, we get
\begin{align}
    \langle \lambda, T_\lambda, W_\lambda | \hat{X}^{(N)} | \nu, T_\nu, W_\nu\rangle & \nonumber \\
    & \hspace{-4cm} = \sum_{i,j = 0}^{d-1}
    \sum_{\mbox{\scriptsize $\begin{array}{c} W_{\nu(N-1)} \\ \in \mcal{W}_{\nu(N-1)} \end{array}$}}
    \left( \begin{array}{ccc} \lambda & \nu(N-1) & \nu \\ W_{\lambda} & W_{\nu(N-1)} & W_{\nu} \end{array} \right)_{i,j} \langle i | \hat{X} | j\rangle \delta_{\lambda(N-1), \nu(N-1)} \delta_{T_{\lambda(N-1)}, T_{\nu(N-1)}} \nonumber \\
& \hspace{-4cm} = \mathrm{Tr}[\hat{g}_{\nu(N-1)}^{(\lambda,W_\lambda;\nu,W_\nu)\dagger} \hat{X}] \delta_{\lambda(N-1), \nu(N-1)} \delta_{T_{\lambda(N-1)}, T_{\nu(N-1)}}.
    \label{el}
\end{align}
Inserting this result into Eq.~(\ref{gen}) and observing that $\sum_{T_\lambda} = \sum_{\lambda^-}\sum_{T_{\lambda^-}}$ and similarly for the sum over $T_\nu$, the general Identity (\ref{genIdentity1}) is directly obtained. Interestingly, Eq.~(\ref{el}) also shows that
\begin{equation}
    \langle \hat{X}^{(N)} \rangle_{|\nu,T_\nu,W_\nu\rangle} = \langle \hat{X} \rangle_{\hat{\rho}_{\nu(N-1)}^{(\nu,W_\nu)}}, \quad \forall \hat{X}.
\end{equation}

\section{Generalization to PI Liouvillians with $p$-particle terms}
\label{generalpPILindApp}

We consider here the generalized case where the PI Liouvillian contains $p$-particle terms, i.e., is of the form
\begin{equation}
\mcal{L} = \sum_p \left(\mcal{V}_{\hat{H}_{p,c}} + \mcal{D}^{(\mathrm{loc})}_{\hat{\ell}_p} + \mcal{D}^{(\mathrm{col})}_{\hat{L}_p}\right),
\end{equation}
where the sum over $p$ only runs for values between 1 and $N$ for which either of the three Liouvillian $p$-particle terms is nonzero, $\hat{H}_{p,c} = \sum_{n_1 < \cdots < n_p = 1}^N \hat{H}_p^{(n_1,\ldots,n_p)}$, $\mcal{D}^{(\mathrm{loc})}_{\hat{\ell}_p} = \gamma_p \sum_{n_1 < \cdots < n_p=1}^N \mcal{D}_{\hat{\ell}_p^{(n_1,\ldots,n_p)}}$, and $\mcal{D}^{(\mathrm{col})}_{\hat{L}_p} = \gamma_{p,c} \mcal{D}_{\hat{L}_{p,c}}$, with $\hat{L}_{p,c} = \sum_{n_1 < \cdots < n_p = 1}^N \hat{L}_p^{(n_1,\ldots,n_p)}$. Here, $\hat{H}_p$ is a $p$-particle Hamiltonian, $\hat{\ell}_p$ and $\hat{L}_p$ are $p$-particle jump operators, and $(n_1,\ldots, n_p)$ denotes the particle $p$-uple these $p$-particle operators act on. These operators satisfy $\hat{X}_p^{(n_1,\ldots,n_p)} = \hat{X}_p^{(n_{\pi(1)},\ldots,n_{\pi(p)})}$, for all $n_1 \neq \ldots \neq n_p$ and permutations $\pi \in S_p$.

All Liouvillian terms can again be expressed with the help of generic superoperators, namely $\mcal{K}_{\hat{X}_p,\hat{Y}_p}$ ($\hat{X}_p, \hat{Y}_p$ any two $p$-particle operators) that act according to
\begin{equation}
    \mcal{K}_{\hat{X}_p,\hat{Y}_p}[\hat{A}] = \sum_{n_1 < \cdots < n_p = 1}^N \hat{X}_p^{(n_1,\ldots,n_p)} \hat{A} \hat{Y}_p^{(n_1,\ldots,n_p)\dagger}, \quad \forall \hat{A} \in \mathcal{L}(\mcal{H}).
\end{equation}
Indeed, $\mcal{V}_{\hat{H}_{p,c}} = (i/\hbar) (\mcal{K}_{\hat{\mathbbm{1}},\hat{H}_p} - \mcal{K}_{\hat{H}_p,\hat{\mathbbm{1}}})$,
$\mcal{D}_{\hat{\ell}_p}^{(\mathrm{loc})} = \gamma_p(\mcal{K}_{\hat{\ell}_p,\hat{\ell}_p} - (\mcal{K}_{\hat{\ell}^\dagger_p \hat{\ell}_p,\hat{\mathbbm{1}}} + \mcal{K}_{\hat{\mathbbm{1}},\hat{\ell}^\dagger_p \hat{\ell}_p})/2)$, and
$\mcal{D}_{\hat{L}_p}^{(\mathrm{col})}[\hat{\rho}] = \gamma_{p,c}(\hat{L}_{p,c} \mcal{K}_{\hat{\mathbbm{1}},\hat{L}_p}[\hat{\rho}] - (\hat{L}_{p,c}^\dagger \mcal{K}_{\hat{L}_p,\hat{\mathbbm{1}}}[\hat{\rho}] + \mcal{K}_{\hat{\mathbbm{1}},\hat{L}_p}[\hat{\rho}] \hat{L}_{p,c})/2)$,
where $\hat{L}_{p,c}$ can similarly be written as $\mcal{K}_{\hat{L}_p,\hat{\mathbbm{1}}}[\hat{\mathbbm{1}}]$. The superoperators $\mcal{K}_{\hat{X}_p,\hat{Y}_p}$ are PI, so that $\mcal{K}_{\hat{X}_p,\hat{Y}_p}[\hat{A}_\mathrm{PI}]$ is itself a PI operator for any PI operator $\hat{A}_\mathrm{PI}$. With respect to Hermitian conjugation, we have $\mcal{K}_{\hat{X}_p,\hat{Y}_p}[\hat{A}]^\dagger = \mcal{K}_{\hat{Y}_p,\hat{X}_p}[\hat{A}^\dagger]$.

To get explicit expressions of the matrix elements $\mcal{L}_{\lambda,W_\lambda,W'_\lambda; \nu, W_\nu,W'_\nu}$, it is therefore again completely enough to
have the expansion in the commutant operator basis of the PI operators $\mcal{K}_{\hat{X}_p,\hat{Y}_p}[\hat{F}_\nu^{(W_\nu,W'_\nu)}]$, $\forall \hat{X}_p,\hat{Y}_p,\nu, W_\nu, W'_\nu$. This expansion reads
\begin{equation}
\mcal{K}_{\hat{X}_p,\hat{Y}_p}[\hat{F}_\nu^{(W_\nu,W'_\nu)}] = \sum_{\lambda \vdash (N,d)} \sum_{W_\lambda, W'_\lambda \in \mcal{W}_\lambda}  \mathrm{Tr}(\hat{F}_\lambda^{(W_\lambda,W'_\lambda)\dagger} \mcal{K}_{\hat{X}_p,\hat{Y}_p}[\hat{F}_\nu^{(W_\nu,W'_\nu)}]) \hat{F}_\lambda^{(W_\lambda,W'_\lambda)}.
\end{equation}

Since $\mcal{P}_\sigma[\hat{X}_p^{(n_1,\ldots,n_p)} \hat{A}_\mathrm{PI} \hat{Y}_p^{(n_1,\ldots,n_p)\dagger}] = \hat{X}_p^{(\sigma(n_1),\ldots,\sigma(n_p))} \hat{A}_\mathrm{PI} \hat{Y}^{(\sigma(n_1),\ldots,\sigma(n_p))\dagger}$, for all $n_1 \neq \cdots \neq n_p$ and permutations $\sigma \in S_N$, $\mathrm{Tr}(\hat{F}_\lambda^{(W_\lambda,W'_\lambda)\dagger} \hat{X}_p^{(n_1,\ldots,n_p)} \hat{A}_\mathrm{PI} \hat{Y}_p^{(n_1,\ldots,n_p)\dagger})$ is independent of the $p$-uple $(n_1,\ldots, n_p)$ and we get
\begin{align}
\mathrm{Tr}(\hat{F}_\lambda^{(W_\lambda,W'_\lambda)\dagger} \mcal{K}_{\hat{X}_p,\hat{Y}_p}[\hat{F}_\nu^{(W_\nu,W'_\nu)}])
& \nonumber \\
& \hspace{- 3 cm} = \binom{N}{p} \mathrm{Tr}(\hat{F}_\lambda^{(W_\lambda,W'_\lambda)\dagger} \hat{X}_p^{(N-p+1,\ldots, N)} \hat{F}_\nu^{(W_\nu,W'_\nu)} \hat{Y}_p^{(N-p+1,\ldots,N)\dagger}) \nonumber \\
& \hspace{- 3 cm} = \binom{N}{p}\frac{1}{\sqrt{f^\lambda f^\nu}} \sum_{T_\lambda, T_\nu} \langle \lambda, T_\lambda, W_\lambda|\hat{X}_p^{(N-p+1,\ldots,N)}|\nu, T_\nu, W_\nu\rangle \nonumber \\
& \qquad \langle \lambda, T_\lambda, W'_\lambda|\hat{Y}_p^{(N-p+1,\ldots,N)}|\nu, T_\nu, W'_\nu\rangle^*.
\label{genp}
\end{align}

Applying successively $p$ times Eq.~(28) first to isolate the $N$-th qudit, then the $(N-1)$-th, and so on until the $(N-p+1)$-th, we get in similar notations
\renewcommand{\arraystretch}{1.1}
\begin{align}
|\nu,T_\nu,W_\nu\rangle & = \sum_{j_p = 0}^{d - 1} \sum_{\mbox{\scriptsize $\begin{array}{c} W_{\nu(N-1)} \\ \in \mcal{W}_{\nu(N-1)} \end{array}$}} \langle W_{\nu(N-1)},j_p | W_\nu \rangle |\nu(N-1),T_{\nu(N-1)},W_{\nu(N-1)}\rangle \otimes |j_p\rangle_N \nonumber \\
& = \sum_{j_{p-1},j_p = 0}^{d - 1}  \sum_{\mbox{\scriptsize $\begin{array}{c} W_{\nu(N-1)} \\ \in \mcal{W}_{\nu(N-1)} \end{array}$}}  \sum_{\mbox{\scriptsize $\begin{array}{c} W_{\nu(N-2)} \\ \in \mcal{W}_{\nu(N-2)} \end{array}$}} \langle W_{\nu(N-1)},j_p | W_\nu \rangle \langle W_{\nu(N-2)},j_{p-1} | W_{\nu(N-1)} \rangle \nonumber \\
& \qquad \qquad \qquad \qquad \qquad \qquad |\nu(N-2), T_{\nu(N-2)},W_{\nu(N-2)}\rangle \otimes |j_{p-1},j_p\rangle_{N-1,N} \nonumber \\
& \quad \vdots \label{nuTnWup} \\
& =  \sum_{j_1,\ldots, j_p = 0}^{d - 1} \sum_{\mbox{\scriptsize $\begin{array}{c} W_{\nu(N-1)} \\ \in \mcal{W}_{\nu(N-1)} \end{array}$}} \ldots \sum_{\mbox{\scriptsize $\begin{array}{c} W_{\nu(N-p)} \\ \in \mcal{W}_{\nu(N-p)} \end{array}$}} \langle W_{\nu(N-1)},j_p | W_\nu \rangle \cdots \langle W_{\nu(N-p)},j_1 | W_{\nu(N-p+1)} \rangle \nonumber \\
& \qquad \qquad \qquad \qquad \qquad \qquad |\nu(N-p),T_{\nu(N-p)},W_{\nu(N-p)} \rangle \otimes |j_1,\ldots, j_p\rangle_{N-p+1,\ldots, N}, \nonumber
\end{align}
\renewcommand{\arraystretch}{1}
with, $\forall k = 1, \ldots, p$, $\nu(N-k)$ the shape of $T_\nu$ without boxes $N, \ldots, N-k+1$, and $T_{\nu(N-k)}$ the SYT $T_\nu$ without these $k$ boxes.

For every $\nu_L, \nu_R \in \mathcal{P}_d$, $W_\mu \in \mcal{W}_\mu$ ($\mu = \nu_L, \nu_R$), $\boldsymbol{\nu} \equiv (\nu_{l,p-1}, \ldots, \nu_{l,1}, \nu_c, \nu_{r,1}, \ldots, \nu_{r,p-1}) \in \mathcal{P}_d^{2p-1}$ (i.e., $\boldsymbol{\nu}$ is a vector of $2p-1$ partitions of at most $d$ parts), $\boldsymbol{W}_{\boldsymbol{\nu}} \equiv (W_{\nu_{l,p-1}}, \ldots, W_{\nu_{l,1}}, W_{\nu_c}, W_{\nu_{r,1}}, \ldots, W_{\nu_{r,p-1}})$, with $W_\mu \in \mcal{W}_\mu$ ($\mu = \nu_{l,p-1}, \ldots, \nu_{r,p-1}$), we define the generalized $3\nu$ symbol \scalebox{0.6}{$\genthreenu{\nu_L}{\nu}{\nu_R}$} as being the square $d^p \times d^p$ matrix with entries
\begin{equation}
    \genthreenu{\nu_L}{\nu}{\nu_R}_{\mathbf{i},\mathbf{j}} = \prod_{k = 1}^p \langle W_{\nu_{l,k-1}},i_k| W_{\nu_{l,k}} \rangle \langle W_{\nu_{r,k-1}},j_k| W_{\nu_{r,k}} \rangle,
\end{equation}
where $\mathbf{i} \equiv (i_1, \ldots, i_p)$, $\mathbf{j} \equiv (j_1,\ldots,j_p)$, $i_k, j_k = 0, \ldots, d-1$ for $k = 1, \ldots, p$, and where we set $\nu_{l,0} = \nu_{r,0} \equiv \nu_c$, $\nu_{l,p} \equiv \nu_L$, and $\nu_{r,p} \equiv \nu_R$.

The generalized $3\nu$-symbol matrix \scalebox{0.6}{$\genthreenu{\nu_L}{\nu}{\nu_R}$} is necessarily zero if the condition $\nu_{l,k-1} \in \{\nu_{l,k}^-\}$ and $\nu_{r,k-1} \in \{\nu_{r,k}^-\}$, $\forall k = 1, \ldots, p$ (\emph{generalized partition triangle selection rule}) is not satisfied. This condition can only be met if $\nu_L \in \{\nu_R^{-^p+^p}\}$ or equivalently $\nu_R \in \{\nu_L^{-^p+^p}\}$, where the superscript $^{-^p}$ [$^{+^p}$] denotes the action of removing [adding] successively $p$ inner [outer] corners to the partition it applies. We define the \emph{generalized partition triangular delta} $\{\nu_L, \boldsymbol{\nu}, \nu_R \}$ to be $1$ if the generalized partition triangle selection rule is satisfied and $0$ otherwise.

Since the CGC's are real, we have
\begin{equation}
\genthreenu{\nu_L}{\nu}{\nu_R} = \genthreenu{\nu_R}{\tilde{\nu}}{\nu_L}^T,
\end{equation}
where $\boldsymbol{\tilde{\nu}}$ is the vector of partitions $\boldsymbol{\nu}$ listed in reversed order: $\boldsymbol{\tilde{\nu}} \equiv (\nu_{r,p-1}, \ldots, \nu_{r,1}, \nu_c, \nu_{l,1}, \ldots, \nu_{l,p-1})$. The generalized $3\nu$-symbol matrices obey the generalized orthogonality relation (see~\ref{CGApp})
\begin{equation}
    \sum_{\boldsymbol{W}_{\boldsymbol{\nu}}} \mathrm{Tr}\left[\genthreenu{\nu_L}{\nu}{\nu_R}\right] = \{\nu_L,\boldsymbol{\nu},\nu_R\} \delta_{\nu_L, \nu_R}\delta_{W_{\nu_L},W_{\nu_R}} \delta_{\boldsymbol{\nu},\boldsymbol{\tilde{\nu}}},
\end{equation}
with $\sum_{\boldsymbol{W}_{\boldsymbol{\nu}}} \equiv \sum_{W_{\nu_{l,p-1}}} \ldots \sum_{W_{\nu_{l,1}}} \sum_{W_{\nu_c}} \sum_{W_{\nu_{r,1}}} \ldots \sum_{W_{\nu_{r,p-1}}}$, and they are the representation matrices in the computational basis of the $p$-qudit product operators
\begin{equation}
\hat{g}_{\boldsymbol{\nu},\boldsymbol{W}_{\boldsymbol{\nu}}}^{(\nu_L,W_{\nu_L};\nu_R,W_{\nu_R})} = \bigotimes_{k = 1}^p |\phi_{\nu_{l,k-1},W_{\nu_{l,k-1}}}^{(\nu_{l,k},W_{\nu_{l,k}})}\rangle \langle \phi_{\nu_{r,k-1},W_{\nu_{r,k-1}}}^{(\nu_{r,k},W_{\nu_{r,k}})}|.
\end{equation}

We also define the $p$-qudit operators
\begin{equation}
    \hat{g}_{\boldsymbol{\nu}}^{(\nu_L,W_{\nu_L};\nu_R,W_{\nu_R})} = \sum_{\boldsymbol{W}_{\boldsymbol{\nu}}} \hat{g}_{\boldsymbol{\nu},\boldsymbol{W}_{\boldsymbol{\nu}}}^{(\nu_L,W_{\nu_L};\nu_R,W_{\nu_R})}.
\end{equation}
These operators vanish if $\{\nu_L,\boldsymbol{\nu},\nu_R\} = 0$ and they satisfy $\hat{g}^{(\nu_L,W_{\nu_L};\nu_R,W_{\nu_R})\dagger}_{\boldsymbol{\nu}} = \hat{g}^{(\nu_R,W_{\nu_R};\nu_L,W_{\nu_L})}_{\boldsymbol{\tilde{\nu}}}$ and
\begin{equation}
\mathrm{Tr}[\hat{g}^{(\nu_L,W_{\nu_L};\nu_R,W_{\nu_R})}_{\boldsymbol{\nu}}] = \{\nu_L,\boldsymbol{\nu},\nu_R\} \delta_{\nu_L, \nu_R}\delta_{W_{\nu_L},W_{\nu_R}} \delta_{\boldsymbol{\nu},\boldsymbol{\tilde{\nu}}}.
\label{genTrq}
\end{equation}
As a result, $\forall \nu \in \mathcal{P}_d$, $W_\nu \in \mcal{W}_\nu$, $\boldsymbol{\mu} \in \mathcal{P}_d^{2p-1}: \boldsymbol{\mu} = \boldsymbol{\tilde{\mu}}$ and $\{\nu,\boldsymbol{\mu},\nu\} = 1$, $\hat{\rho}^{(\nu,W_\nu)}_{\boldsymbol{\mu}} \equiv \hat{g}^{(\nu,W_\nu;\nu,W_\nu)}_{\boldsymbol{\mu}}$ is a trace 1 positive semidefinite operator and represents a separable $p$-qudit mixed state.

With this stated and expanding $|\lambda, T_\lambda, W_\lambda\rangle$ similarly as $|\nu,T_\nu,W_\nu\rangle$ in Eq.~(\ref{nuTnWup}), we directly obtain
\begin{align}
    \langle \lambda, T_\lambda, W_\lambda | \hat{X}_p^{(N-p+1,\ldots,N)} | \nu, T_\nu, W_\nu\rangle & \nonumber \\
    & \hspace{-5 cm} = \sum_{\mathbf{i},\mathbf{j}}
    \sum_{\boldsymbol{W}_{\boldsymbol{\mu}_{(T_\lambda, T_\nu)_p}}}
    \left( \begin{array}{ccc} \lambda & \boldsymbol{\mu}_{(T_\lambda,T_\nu)_p} & \nu \\ W_\lambda & \boldsymbol{W}_{\boldsymbol{\mu}_{(T_\lambda,T_\nu)_p}} & W_{\nu} \end{array} \right)_{\mathbf{i},\mathbf{j}} \langle \mathbf{i} | \hat{X}_p | \mathbf{j} \rangle \delta_{\lambda(N-p), \nu(N-p)} \delta_{T_{\lambda(N-p)}, T_{\nu(N-p)}} \nonumber \\
& \hspace{-5 cm} = \mathrm{Tr}[\hat{g}_{\boldsymbol{\mu}_{(T_\lambda, T_\nu)_p}}^{(\lambda,W_\lambda;\nu,W_\nu)\dagger} \hat{X}_p] \delta_{\lambda(N-p), \nu(N-p)} \delta_{T_{\lambda(N-p)}, T_{\nu(N-p)}},
    \label{elp}
\end{align}
with $\boldsymbol{\mu}_{(T_\lambda,T_\nu)_p} = (\lambda(N-1), \ldots, \lambda(N-p+1), \lambda(N-p), \nu(N-p+1), \ldots, \nu(N-1))$. Interestingly, this also implies that
\begin{equation}
    \langle \hat{X}_p^{(N-p+1,\ldots,N)}\rangle_{|\nu, T_\nu, W_\nu\rangle} = \langle \hat{X}_p\rangle_{\hat{\rho}^{(\nu,W_\nu)}_{\boldsymbol{\mu}_{(T_\nu)_p}}}, \quad \forall \hat{X}_p,
\end{equation}
with $\boldsymbol{\mu}_{(T_\nu)_p} \equiv \boldsymbol{\mu}_{(T_\nu,T_\nu)_p}$.

Inserting Eq.~(\ref{elp}) into Eq.~(\ref{genp}) and observing that $\sum_{T_\lambda} = \sum_{\lambda^-}\sum_{T_{\lambda^-}} = \sum_{\lambda^-} \sum_{(\lambda^-)^-}\sum_{T_{(\lambda^-)^-}} = \cdots $ (so on $p$ times) and similarly for the sum over $T_\nu$, we immediately get
\begin{equation}
\mcal{K}_{\hat{X}_p,\hat{Y}_p}[\hat{F}_\nu^{(W_\nu,W'_\nu)}] = \sum_{\lambda \in \{\nu^{-^p +^p}\}} \sum_{W_\lambda, W'_\lambda \in \mcal{W}_\lambda}  K_{\hat{X}_p,\hat{Y}_p}^{(\lambda,W_\lambda,W'_\lambda;\nu,W_\nu,W'_\nu)} \hat{F}_\lambda^{(W_\lambda,W'_\lambda)},
\label{genIdentityp}
\end{equation}
with
\begin{equation}
K_{\hat{X}_p,\hat{Y}_p}^{(\lambda,W_\lambda,W'_\lambda;\nu,W_\nu,W'_\nu)} = \sum_{\mbox{\scriptsize $\begin{array}{c}\boldsymbol{\mu} \in \mathcal{P}_d^{2p-1}: \\ \{\lambda,\boldsymbol{\mu},\nu\} = 1 \end{array}$}} \sqrt{r^{\mu_c}_\lambda r^{\mu_c}_\nu} \,
\mathrm{Tr}[\hat{g}_{\boldsymbol{\mu}}^{(\lambda,W_\lambda;\nu,W_\nu)\dagger} \hat{X}_p] \, \mathrm{Tr}[\hat{g}_{\boldsymbol{\mu}}^{(\lambda,W'_\lambda;\nu,W'_\nu)\dagger} \hat{Y}_p]^\ast,
\label{idp}
\end{equation}
where we defined $r^{\mu}_\nu \equiv \binom{N}{p} f^{\mu}/f^\nu$, $\forall \nu \vdash N, \mu \in \{\nu^{-^p}\}$. Identity~(\ref{genIdentityp}) states that the matrix elements of the superoperator 
$\mcal{K}_{\hat{X}_p,\hat{Y}_p}$ are merely given by
\begin{equation}
[\mcal{K}_{\hat{X}_p,\hat{Y}_p}]_{\lambda, W_\lambda, W'_\lambda; \nu, W_\nu, W'_\nu} = K_{\hat{X}_p,\hat{Y}_p}^{(\lambda,W_\lambda,W'_\lambda;\nu,W_\nu,W'_\nu)} \delta_{\lambda,\{\nu^{-^p+^p}\}},
\end{equation}
where we have added here the factor $\delta_{\lambda,\{\nu^{-^p+^p}\}}$ (1 if $\lambda \in \{\nu^{-^p+^p}\}$ and 0 otherwise) for an explicit reference on when the matrix elements are necessarily zero or not (this factor is superfluous since the generalized partition triangle selection rule discussed above implies $K_{\hat{X},\hat{Y}}^{(\lambda, W_\lambda, W'_\lambda; \nu, W_\nu, W'_\nu)} = 0$ if $\lambda \notin \{\nu^{-^p+^p}\}$).

Thanks to Eq.~(\ref{genTrq}), the coefficients $K_{\hat{X}_p,\hat{\mathbbm{1}}_p}^{(\lambda, W_\lambda, W'_\lambda; \lambda, \tilde{W}_\lambda, W'_\lambda)}$, with $\hat{\mathbbm{1}}_p$ the $p$-particle identity, are independent of $W'_\lambda$ and we can define
\begin{equation}
K_{\hat{X}_p}^{(\lambda,W_\lambda, \tilde{W}_\lambda)} \equiv K_{\hat{X}_p,\hat{\mathbbm{1}}_p}^{(\lambda, W_\lambda, W'_\lambda; \lambda, \tilde{W}_\lambda, W'_\lambda)} = \sum_{\mbox{\scriptsize $\begin{array}{c} \boldsymbol{\mu} \in \mathcal{P}_d^{2p-1} : \\ \{\lambda,\boldsymbol{\mu},\lambda \} = 1 \, \&\, \boldsymbol{\mu} = \boldsymbol{\tilde{\mu}} \end{array}$}} r^{\mu_c}_\lambda \, \mathrm{Tr}[\hat{g}_{\boldsymbol{\mu}}^{(\lambda,W_\lambda;\lambda,\tilde{W}_\lambda)\dagger} \hat{X}_p].
\end{equation}
This yields
\begin{equation}
K_{\hat{X}_p,\hat{\mathbbm{1}}_p}^{(\lambda,W_\lambda,W'_\lambda;\nu,W_\nu,W'_\nu)} = K_{\hat{X}_p}^{(\lambda, W_\lambda, W_\nu)} \delta_{\lambda, \nu} \delta_{W'_\lambda, W'_\nu}
\end{equation}
and subsequently $\mcal{K}_{\hat{X}_p,\hat{\mathbbm{1}}_p}[\hat{F}_\nu^{(W_\nu,W'_\nu)}] = \sum_{\tilde{W}_\nu} K_{\hat{X}_p}^{(\nu, \tilde{W}_\nu, W_\nu)} \hat{F}_\nu^{(\tilde{W}_\nu,W'_\nu)}$ and $\hat{X}_{p,c} = \sum_{\nu, W_\nu, W'_\nu} \sqrt{f^\nu} K_{\hat{X}_p}^{(\nu,W_\nu, W'_\nu)} \hat{F}_\nu^{(W_\nu,W'_\nu)}$, so that the master equation matrix elements merely generalizes in presence of $p$-particle operators according to
\begin{align}
[\mcal{V}_{\hat{H}_{p,c}}]_{\lambda, W_\lambda, W'_\lambda; \nu, W_\nu, W'_\nu} & = \frac{i}{\hbar} \left( K_{\hat{H}_p}^{(\nu, W'_\nu,W'_\lambda)} \delta_{W_\lambda,W_\nu} - K_{\hat{H}_p}^{(\nu, W_\lambda, W_\nu)} \delta_{W'_\lambda,W'_\nu} \right) \delta_{\lambda, \nu}, \\
[\mcal{D}^{(\mathrm{loc})}_{\hat{\ell}_p}]_{\lambda, W_\lambda, W'_\lambda; \nu, W_\nu, W'_\nu} & = \gamma_p \bigg[ K_{\hat{\ell}_p,\hat{\ell}_p}^{(\lambda, W_\lambda, W'_\lambda; \nu, W_\nu, W'_\nu)} \delta_{\lambda,\{\nu^{-^p+^p}\}} \\
& \qquad \qquad - \frac{1}{2} \bigg( K_{\hat{\ell}_p^\dagger \hat{\ell}_p}^{(\nu, W'_\nu, W'_\lambda)} \delta_{W_\lambda,W_\nu} + K_{\hat{\ell}_p^\dagger \hat{\ell}_p}^{(\nu, W_\lambda, W_\nu)} \delta_{W'_\lambda,W'_\nu}\bigg) \delta_{\lambda, \nu} \bigg] , \nonumber \\
[\mcal{D}^{(\mathrm{col})}_{\hat{L}_p}]_{\lambda, W_\lambda, W'_\lambda; \nu, W_\nu, W'_\nu} & = \gamma_{p,c} \bigg[ K_{\hat{L}_p}^{(\nu, W_\lambda, W_\nu)} K_{\hat{L}_p}^{(\nu, W'_\lambda, W'_\nu)\ast} \nonumber \\ & \qquad \qquad - \frac{1}{2} \bigg( \sum_{\tilde{W}'_\nu} K_{\hat{L}_p}^{(\nu, \tilde{W}'_\nu, W'_\lambda)} K_{\hat{L}_p}^{(\nu, \tilde{W}'_\nu, W'_\nu)\ast} \bigg) \delta_{W_\lambda,W_\nu} \\
& \qquad \qquad - \frac{1}{2} \bigg( \sum_{\tilde{W}_\nu} K_{\hat{L}_p}^{(\nu, \tilde{W}_\nu, W_\nu)} K_{\hat{L}_p}^{(\nu, \tilde{W}_\nu, W_\lambda)\ast} \bigg) \delta_{W'_\lambda,W'_\nu} \bigg] \delta_{\lambda,\nu}.
\nonumber 
\end{align}

For $p=1$, all results of this Appendix just particularize to the standard formalism of the main manuscript.


\begin{thebibliography}{100}
\bibitem{footnote00} The individual qudit basis states are here denoted by $|0\rangle, \ldots, |d-1\rangle$.
\bibitem{footnote0} The action of the permutation operator $\hat{P}_\sigma$ on the computational basis states $|i_1, \ldots, i_N\rangle$ ($i_k = 0, \ldots, d-1$ for $k = 1, \ldots, N$) is standardly defined by $\hat{P}_\sigma |i_1, \ldots, i_N\rangle = |i_{\sigma^{-1}(1)}, \ldots, i_{\sigma^{-1}(N)}\rangle$.
\bibitem{footnote1} The product of two PI operators is PI and so is the Hermitian conjugate of a PI operator: the commutant $\mathcal{L}_{S_N}(\mcal{H})$ is a $\ast$-algebra of operators on $\mcal{H}$.
\bibitem{footnote1b} As a reminder, the Hermitian conjugate of a superoperator $\mcal{L}$ is the unique superoperator $\mcal{L}^\dagger$ such that $\mathrm{Tr}(\hat{A}^\dagger \mcal{L}^\dagger[\hat{B}]) = \mathrm{Tr}(\mcal{L}[\hat{A}]^\dagger \hat{B})$, for all $\hat{A},\hat{B} \in \mathcal{L}(\mcal{H})$.
\bibitem{footnote1c} Within the formalism of the superoperators of permutation $\mcal{P}_\sigma$, a PI operator $\hat{A}_\mathrm{PI}$ is an operator that satisfies $\mcal{P}_\sigma[\hat{A}_\mathrm{PI}] = \hat{A}_\mathrm{PI}, \forall \sigma$.
\bibitem{footnote2} This means that each superoperator $\mcal{P}_\sigma$ defines a $\ast$-isomorphism on the Liouville space $\mathcal{L}(\mcal{H})$.
\bibitem{footnote3} Indeed, $\forall \sigma, n$, and local operator $\hat{X}$, we have $\mcal{P}_\sigma[\hat{X}^{(n)}] = \hat{X}^{(\sigma(n))}$, which implies $\{\mcal{P_\sigma}[\hat{X}^{(n)}]\} = \{\hat{X}^{(\sigma(n))}\} = \{\hat{X}^{(n)}\}$ and $\mcal{P_\sigma}[\hat{X}_c] = \hat{X}_c$.
\bibitem{footnote3b} The commutant $\mathcal{L}_{S_N}(\mcal{H})$ is the subspace of PI operators, i.e., of operators $\hat{A}_\mathrm{PI}$ that satisfy $\mcal{P}_\sigma[\hat{A}_\mathrm{PI}] = \hat{A}_\mathrm{PI}, \forall \sigma$~\cite{footnote1c}. It is therefore nothing but the symmetric subspace of $(\mathcal{L}(\mcal{H}_d))^{\otimes N} = \mathcal{L}(\mcal{H}_d^{\otimes N}) = \mathcal{L}(\mcal{H})$.
\bibitem{footnotefnu} The dimension of the irreducible representation $\mcal{S}^\nu$ is given by the elegant hook length formula: $f^\nu = N!/\prod_{(i,j)\in\nu} h_{(i,j)}$, where the product runs over the hook length $h_{(i,j)}$ of each box $(i,j)$ of the diagram $\nu$~\cite{Frame54}. The historical Frobenius-Young determinantal formula~\cite{Fro, You} can also be used instead: for $\nu \equiv (\nu_1, \ldots, \nu_l)$, $f^\nu = N! \prod_{i < j}(l_i - l_j)/\prod_i l_i!$ with $l_i = \nu_i + l(\nu) - i$ and $l(\nu)$ the length of partition $\nu$. In particular, for $\nu = (N)$, $f^\nu = 1$, and for $\nu \equiv (\nu_1,\nu_2) \vdash (N,2)$, $f^\nu = (\nu_1 - \nu_2 + 1)\binom{N}{\nu_2}/(\nu_1 + 1)$. This also allows one to express the ratio~\cite{footnotenumtau} $r^{\nu^{-_\tau}}_\nu \equiv N f^{\nu^{-_\tau}}/f^\nu$ in the form $r^{\nu^{-_\tau}}_\nu = l_\tau \prod_{i \neq \tau}(l_i - l_\tau + 1)/(l_i - l_\tau)$.
\bibitem{footnotefnud} The dimension of the irreducible representation $\mcal{U}^\nu(d)$ reads $f^\nu(d) = \prod_{1 \leq i < j \leq d} (\nu_i - \nu_j + j - i)/(j - i)$, with $\nu_i \equiv 0, \forall i > l(\nu_i)$ (Weyl dimension formula). In particular, $f^{(.)}(d) = 1$ [$(.)$ is the empty partition of length 0 and $\mcal{U}^{(.)}(d)$ is the trivial representation], $f^{(1)}(d) = d$, and $f^{(N)}(d) = \binom{N + d - 1}{N}, \forall N > 0$. The number $f^\nu(d)$ can also be expressed in the form $f^\nu(d) = s_\nu(1, \ldots, 1)$, where $(1, \ldots, 1)$ is a $d$-uple and $s_\nu(x_1, \ldots, x_d)$ is the Schur's polynomial in the $d$ variables $x_1$, \ldots, $x_d$ associated to partition $\nu$ (an homogenous symmetric polynomial of degree $|\nu|$, with $|\nu|$ the weight of $\nu$)~\cite{Okounkov}.
\bibitem{footnoteTnuWnud} The irreps of the subgroups $S_i$ ($1 \leq i < N$) a Schur basis vector $|\nu, T_\nu, W_\nu\rangle$ belongs to are given by the shapes of the SYT $T_\nu$ restricted to only boxes $1$ to $i$. Similarly, the irreps of the subgroups $U(k)$ ($1 \leq k < d$) the vector $|\nu, T_\nu, W_\nu\rangle$ belongs to are given by the shapes of the SWT $W_\nu$ restricted to only boxes $0$ to $k-1$.
\bibitem{footnote4b} Hence, replacing $\hat{F}_\nu^{(W_\nu,W'_\nu)}$ and $\hat{F}_\nu^{(W'_\nu,W_\nu)}$ by $(\hat{F}_\nu^{(W_\nu,W'_\nu)} + \hat{F}_\nu^{(W'_\nu,W_\nu)})/\sqrt{2}$ and $i(\hat{F}_\nu^{(W_\nu,W'_\nu)} - \hat{F}_\nu^{(W'_\nu,W_\nu)})/\sqrt{2}$, $\forall \nu \vdash (N,d)$, $W_\nu, W'_\nu \in \mcal{W}_\nu: W_\nu \neq W'_\nu$, yields together with the operators $\hat{F}_\nu^{(W_\nu,W_\nu)}$ an orthonormal basis of PI Hermitian operators in the commutant $\mathcal{L}_{S_N}(\mcal{H})$. In addition, having $\mathrm{Tr}[\hat{F}_\nu^{(W_\nu,W'_\nu)}] = \sqrt{f^\nu}\delta_{W_\nu,W'_\nu}$, an orthogonal basis in the subspace of traceless PI Hermitian operators is straightfowardly obtained by further replacing all but one operators $\hat{F}_\nu^{(W_\nu,W_\nu)}$ by traceless linear combinations of them.
\bibitem{footnote6} The product of two $\nu$-type operators is a $\nu$-type operator and so is the Hermitian conjugate of a $\nu$-type operator: each operator subspace $\mathcal{L}_\nu(\mcal{H})$ is a $\ast$-algebra of operators on $\mcal{H}$ and a subalgebra of the commutant $\mathcal{L}_{S_N}(\mcal{H})$.
\bibitem{footnoteCorner} The diagram or shape of a partition $\nu \equiv (\nu_1, \ldots, \nu_l) \vdash N$ is an array of $N$ boxes arranged on $l$ left-justified rows, with row $i$ ($1 \leq i \leq l$) containing $\nu_i$ boxes. The shape of a partition $\nu$ is usually denoted by the same symbol $\nu$. An inner corner of a shape $\nu$ is a box $\in \nu$ whose removal leaves us with a valid partition shape. An outer corner of $\nu$ is a box $\notin \nu$ whose addition produces a valid partition shape.
\bibitem{footnote7}
Having $\sum_{\lambda^-} f^{\lambda^-} = f^\lambda$, we get in particular $K_{\hat{\mathbbm{1}}}^{(\lambda,W_\lambda, \tilde{W}_\lambda)}$ $= N \delta_{W_\lambda,\tilde{W}_\lambda}$, so that Eq.~(\ref{genidentity1X1}) yields as expected from definition $\mcal{K}_{\hat{\mathbbm{1}},\hat{\mathbbm{1}}}[\hat{F}_\nu^{(W_\nu,W'_\nu)}] = N \hat{F}_\nu^{(W_\nu,W'_\nu)}$. A similar expression as Eq.~(\ref{genidentity1X1}) is also directly obtained for the operator $\mcal{K}_{\hat{\mathbbm{1}},\hat{Y}}[\hat{F}_\nu^{(W_\nu,W'_\nu)}]$ using the equality $\mcal{K}_{\hat{\mathbbm{1}},\hat{Y}}[\hat{F}_\nu^{(W_\nu,W'_\nu)}] = \mcal{K}_{\hat{Y},\hat{\mathbbm{1}}}[\hat{F}_\nu^{(W'_\nu,W_\nu)}]^\dagger$.
\bibitem{footnotea3} For any operator $\hat{L}$, $\mcal{D}_{\hat{L}}^\dagger[\hat{\rho}] = \hat{L}^\dagger \hat{\rho} \hat{L} - \frac{1}{2} \hat{L}^\dagger \hat{L} \hat{\rho} - \frac{1}{2} \hat{\rho} \hat{L}^\dagger \hat{L}$, so that if $\hat{L}$ is Hermitian, then $\mcal{D}_{\hat{L}}^\dagger = \mcal{D}_{\hat{L}}$.
\bibitem{footnotenumtau} For any partition $\nu$, $\nu^{\mp_\tau}$ denotes the partition obtained by the removal [addition] of the inner [outer] corner of $\nu$ at row $\tau$. We have $\mu = \nu^{\pm_\tau} \Leftrightarrow \nu = \mu^{\mp_\tau}$, so that $\mu \in \{\nu^\pm\} \Leftrightarrow \nu \in \{\mu^\mp\}$.
\bibitem{footnotetaunumu} For any partition $\nu^{\mp}$, the row at which the removal [addition] of the inner [outer] corner of $\nu$ occurs is denoted by $\tau_{\nu^-/\nu}$ [$\tau_{\nu^+/\nu}$]. We have $\tau_{\nu^\pm/\nu} = \tau_{\nu/\nu^\pm}$.
\end{thebibliography}
\end{document}